\documentclass[prb,aps,twocolumn,amsmath,amssymb,showpacs]{revtex4}
\usepackage{epsfig}
\psfigdriver{dvips}
\usepackage{graphicx}
\usepackage{color}
\usepackage{dcolumn} 
\usepackage{bm, bbm}      
\usepackage{curves}
\usepackage{epic}
\usepackage{wasysym}
\usepackage{subfigure}

\def\beq{\begin{equation}}
\def\eeq{\end{equation}}
\def\bea{\begin{eqnarray}}
\def\eea{\end{eqnarray}}

\def\nn{\nonumber}

\begin{document}
\title{
Effects of spin vacancies on magnetic properties of the Kitaev-Heisenberg model}
\author{Fabien Trousselet, Giniyat Khaliullin, and Peter Horsch}
\affiliation{
Max-Planck-Institut f\"ur Festk\"orperforschung, Heisenbergstrasse 1,
D-70569 Stuttgart, Germany}
\date{\today}

\begin{abstract}
We study the ground state properties of the Kitaev-Heisenberg model in a
magnetic field and explore the evolution of spin correlations in the
presence of non-magnetic vacancies. By means of exact diagonalizations, the 
phase diagram without vacancies is determined as a function of the magnetic 
field and the ratio between Kitaev and Heisenberg interactions. We show that 
in the (antiferromagnetic) stripe ordered phase the static susceptibility
and its anisotropy can be described by a spin canting mechanism. This accounts 
as well for the transition to the polarized phase when including quantum 
fluctuations perturbatively. Effects of spin vacancies depend sensitively on 
the type of the ground state. In the liquid phase, the magnetization pattern 
around a single vacancy in a small field is determined, and its spatial 
anisotropy is related to that of \textit{non-zero} further neighbor 
correlations induced by the field and/or Heisenberg interactions.
In the stripe phase, the joint effect of a vacancy and a small field breaks the
six-fold symmetry of the model and stabilizes a particular stripe pattern.
Similar symmetry-breaking effects occur even at zero field due to effective 
interactions between vacancies. This selection mechanism and intrinsic 
randomness of vacancy positions may lead to spin-glass behavior.
\end{abstract}
\maketitle

\section{Introduction}
Much attention has been paid recently to the existence and properties of spin
liquids in strongly correlated materials, both experimentally and 
theoretically\cite{Lee10,Bal10}. Such phases, first proposed by Anderson, 
have been put forward in the context of high $T_c$ superconductivity as a 
possible precursor state, but they can occur in a much broader class of 
situations, both in real materials \cite{SL1,SL2,SL3,Yam08} and in microscopic 
models \cite{trii,tria,trid,hubhex}. Their characterization is challenging --
on the experimental side, it requires generally involved low-temperature 
techniques to prove the absence of long-range magnetic order down 
to temperature $T=0$, and concerning the theory, either 
in Hubbard-like models or in localized spin models, spin liquids usually
result from frustrating interactions but compete with a
variety of possible ordered phases, from usual antiferromagnetic (AF) order to
valence bond crystals, spin nematics etc.\cite{Bal10}

Clear evidence of a spin liquid ground state was found from an exact
solution of the Kitaev model \cite{Kit} of spins $1/2$ on the honeycomb 
lattice. This liquid ground state is characterized by a gapless spectrum (with
low-energy excitations described in terms of Majorana fermions) and gapped
vortices. The Kitaev model itself triggered a large variety of studies,
focusing on topological properties, (non)-abelian excitations, extensions to
higher dimensions\cite{MS}, effects of magnetic\cite{DST} or non-magnetic
impurities and of a magnetic field \cite{Kit,Bask,Wil,Tik}.

Besides its mathematical beauty and the original properties mentioned above, the
Kitaev model was recently found \cite{JK} to be relevant to orbitally 
degenerate systems with strong spin-orbit coupling such as layered iridates 
Na$_2$IrO$_3$ and Li$_2$IrO$_3$. There, around each Ir atom, the octahedral 
environment [see Fig.~\ref{struc}(a)] of oxygens results in a configuration 
with one hole in the $t_{2g}$ manifold; and the large
spin-orbit coupling selects locally a low-energy doublet of states defining a
pseudospin $1/2$; and finally, superexchange processes between Ir ions
driven by moderately strong Coulomb interactions lead to an effective 
description by the Kitaev-Heisenberg model \cite{JK,CJK} for pseudospins 
$1/2$ residing on the honeycomb lattice.
The study of this model showed the robustness of the spin liquid
phase for finite AF Heisenberg interactions [small compared to ferromagnetic
(FM) Kitaev interactions], and a quantum phase transition between this 
liquid phase and a stripe-ordered phase. In Na$_2$IrO$_3$, where interaction 
parameters could be close to this transition, AF order was found by 
magnetization and specific heat measurements\cite{Tak,SG} and x-ray magnetic 
scattering\cite{Liu11}. 

The question of how vacancies influence the order and the magnetic
response in the Kitaev-Heisenberg model is of both experimental and
theoretical interest. Indeed the vacancy can act as a probe allowing to
measure spin correlations in its vicinity; and it can also have
drastic effects on these correlations. For concreteness, 
in $SU(2)$ spin-$1/2$ antiferromagnets with long-range order, a non-magnetic 
impurity can enhance the staggered magnetization around the vacancy 
\cite{egg1,abgh}. In dimerized phases,
the impurity is accompanied by a spinon, which is typically confined to the
impurity in Valence Bond Crystals and deconfined in Resonating Valence Bond
liquids \cite{LhMi}, although the question of the confinement length can be
subtle \cite{PRDV}. In a N\'eel-ordered phase, the impurity results in a 
Curie- type contribution to the finite-$T$ susceptibility \cite{SBV}. 
In the spin-liquid phase of the Kitaev model, a magnetic moment is induced 
around a spinless vacancy; interestingly, its contribution to the spin 
susceptibility diverges logarithmically for $T \rightarrow 0$ \cite{Wil}. 
This makes it relevant to investigate the vacancy-induced magnetic response in 
a model interpolating between $SU(2)$ and Kitaev-like interactions.
Moreover, in the iridate samples studied up to now, a substantial 
amount of disorder might be present, e.g., site-mixing effects with 
non-magnetic alkaline ions on the hexagonal Ir-sublattice \cite{SG}, which 
should be taken into account for proper understanding of their physical 
properties. 

Our aim is to study the effect of non-magnetic vacancies (and of pairs of 
vacancies) in the different magnetic phases of the Kitaev-Heisenberg model, 
at zero temperature. We adopt similar notations as in 
Ref.~\onlinecite{CJK}, with a parameter $0 \leq \alpha \leq 1$ 
interpolating between Heisenberg and Kitaev models, and a uniform magnetic
field $\vec{h}=(h_x,h_y,h_z)$ with amplitude $h=\sqrt{h_x^2+h_y^2+h_z^2}$:
\beq
H= -2\alpha \sum_{\langle i,j \rangle_\gamma} 
\sigma_i^\gamma \sigma_j^\gamma +(1-\alpha)
\sum_{\langle i,j \rangle} \vec{\sigma}_i \cdot \vec{\sigma}_j
-\sum_{i} \vec h \cdot \vec{\sigma}_i ,
\label{ham}
\eeq
where $\gamma=x,y,z$ labels simultaneously an axis in spin space, and 
(in the interaction terms) a bond direction of the honeycomb lattice. 
For convenience we express $H$ in terms of Pauli matrices $\vec{\sigma}_i$.
We take as unit cell a \textit{$z$-bond} [see Fig.~\ref{struc}(b,c)],
on which Kitaev interactions are of the form $\sigma_i^z \sigma_j^z$. Each
site is labeled by an index $i$, in bijection with $(\vec r,\beta)$ where
$\vec r$ denotes the position of the unit cell and $\beta=A,B$ the 
sublattice index, respectively. Elementary translations between neighbor cells
are $\vec{n}_{1/2}=(\pm 1/2,\sqrt{3}/2)$ in Cartesian coordinates.
We consider periodic clusters of $N=24$ sites (this cluster can be represented
with an hexagonal shape, see Fig.~\ref{cormag}), $N=32$ or $N=16$ sites 
(both with the shape of a parallelogram). These clusters respect all spatial
symmetries of the model, i.e. translation and $\pi$-rotation - other
symmetries combining spin- and spatial rotations \cite{JGQT} are not used here. 

In Section \ref{vfs}, we discuss several features of the model in a magnetic
field. We will show how a spin canting mechanism allows one to understand the
phase diagram for a field oriented along an easy axis, as well as to understand  
a directional anisotropy of the static susceptibility. Also the perturbations 
due to Heisenberg interactions and due to small magnetic fields on the spin 
correlation functions are explored. In Section \ref{ces}, we study the effect 
of a single non-magnetic vacancy in the system, in various phases of the 
model, and
analyze the magnetization pattern in the vicinity of this vacancy in a small
field. In Section \ref{ibv}, we discuss situations where two vacancies, close
enough to each other, effectively interact. We focus mainly on the stripe
phase and explain the underlying mechanism of local selection of an ordered
pattern by a vacancy pair. Section \ref{scr} provides a short summary and 
some concluding discussions.

\section{\label{vfs}Vacancy-free system: spin correlations,
  magnetization, phase diagram}
We first consider some basic properties of the Kitaev-Heisenberg model in a
magnetic field, and study the magnetic response to a field 
oriented along different crystallographic directions. This comparison
is important because of the symmetry of the model, which is clearly not $SU(2)$
invariant due to the Kitaev interactions. In the context of iridates, this
anisotropy is due to spin-orbit coupling which results in the easy spin axes 
$x\parallel[100]$, $y\parallel[010]$, $z\parallel[001]$ [corresponding to the 
octahedral axes, see Fig.~\ref{struc}(a)], compared to the cubic diagonal
$[111]$ direction. The clearest anisotropic features are seen in the 
stripe-ordered phase, and we compare this situation to that of the N\'eel
phase.  

\subsection{\label{phas}The competing phases in a magnetic field}

\begin{figure}
\begin{center}
\includegraphics[width=3.5cm]{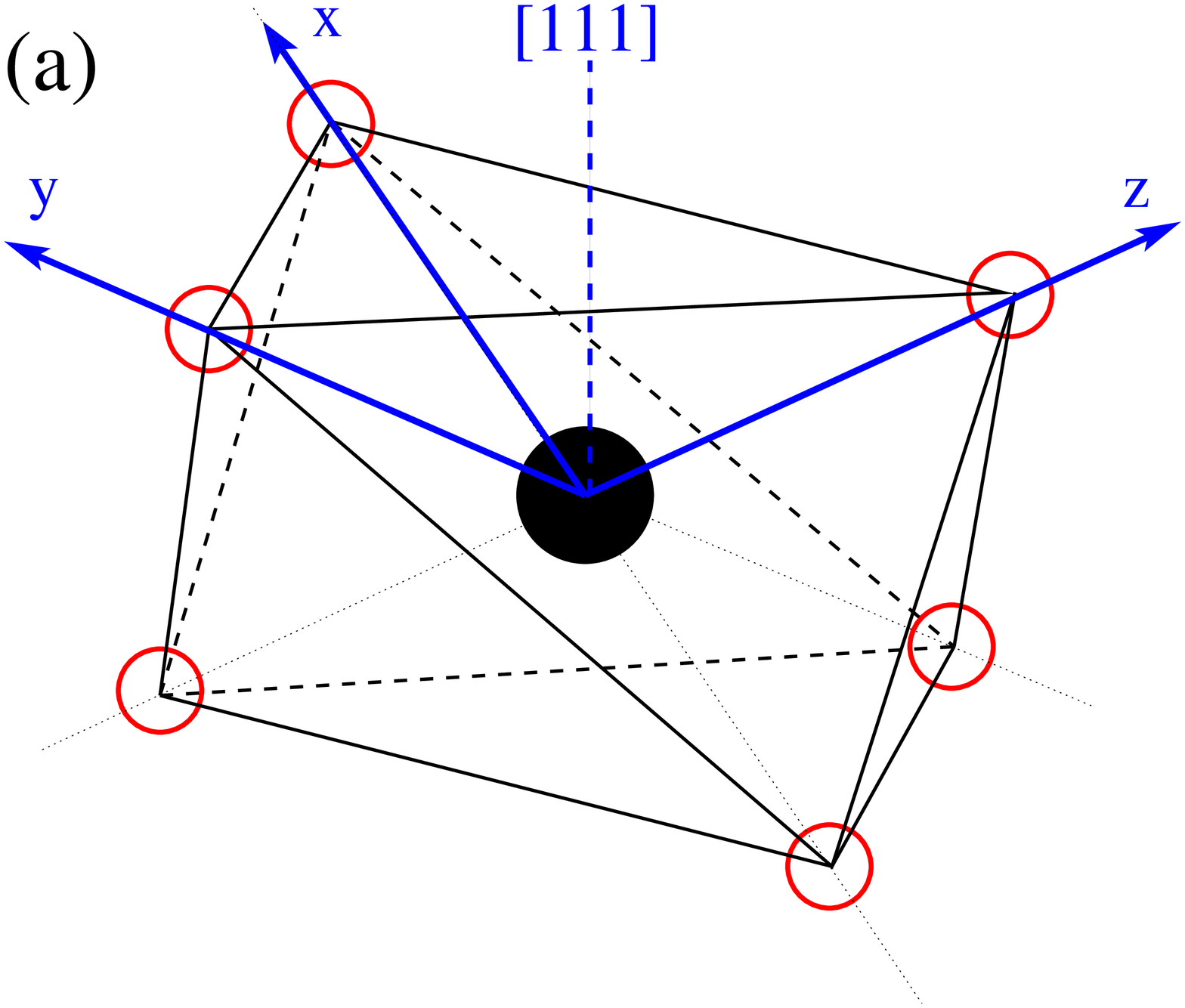}
\includegraphics[width=3.5cm]{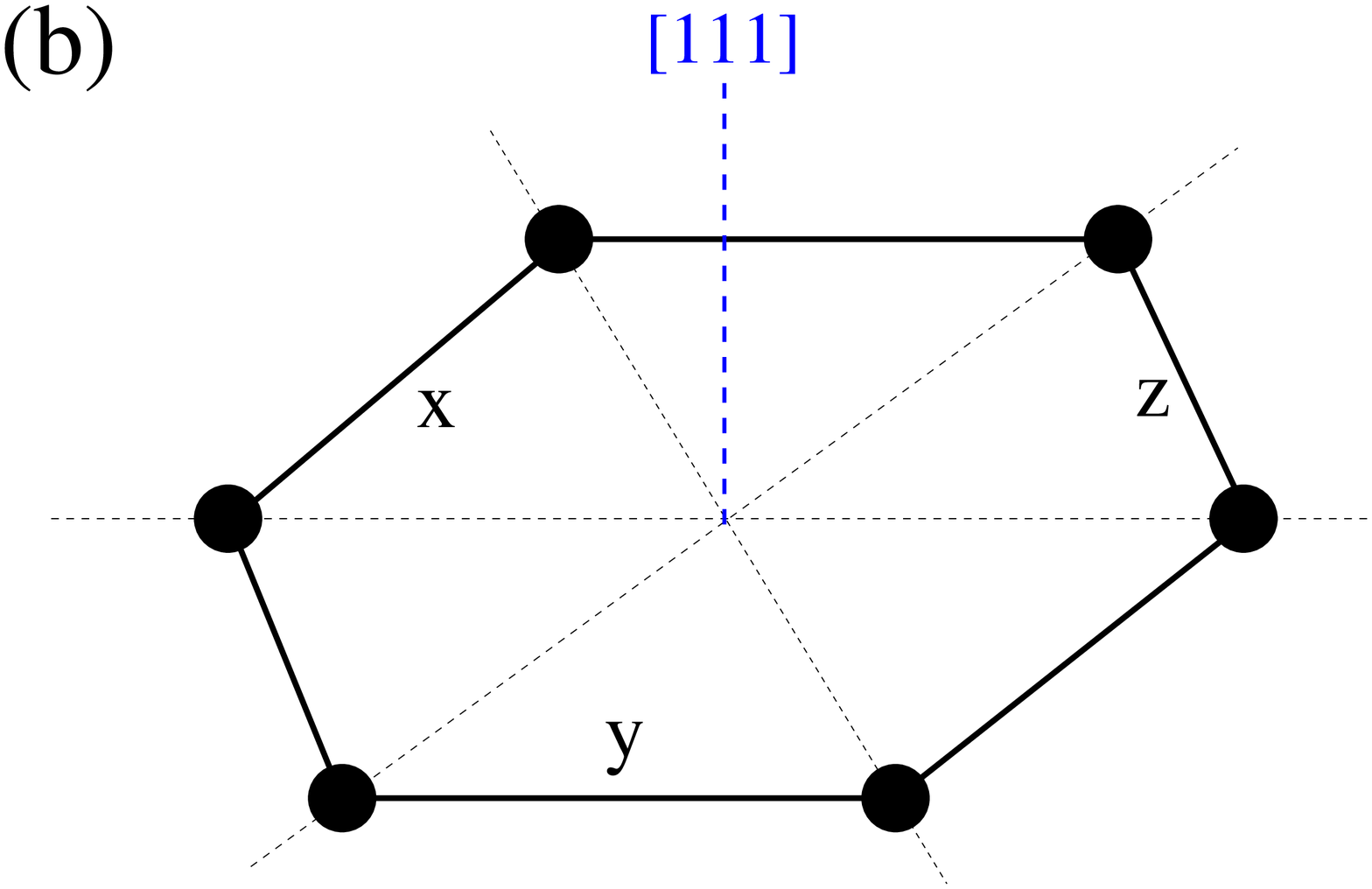}\\
\includegraphics[width=3.6cm]{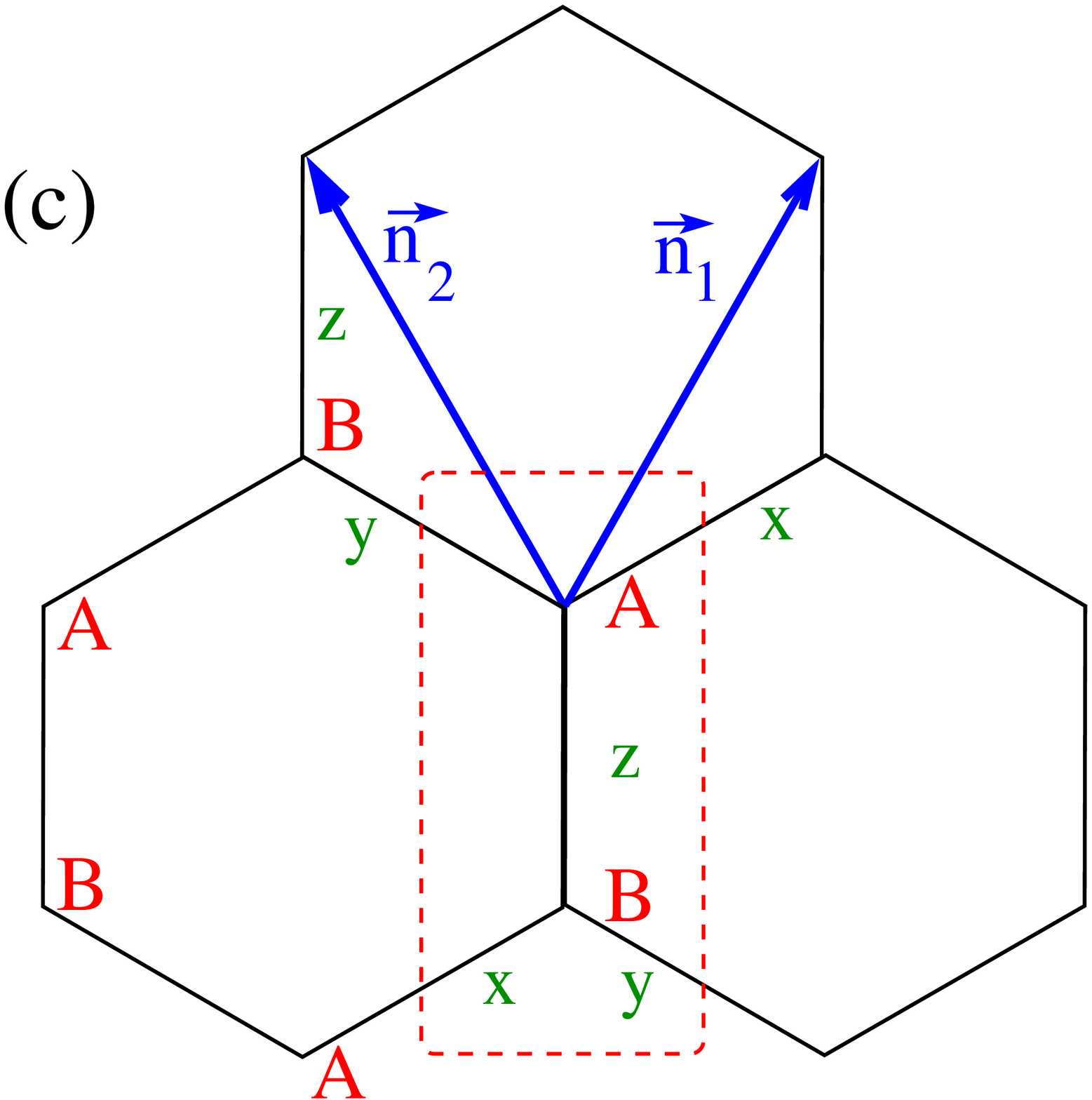}
\includegraphics[width=3.6cm]{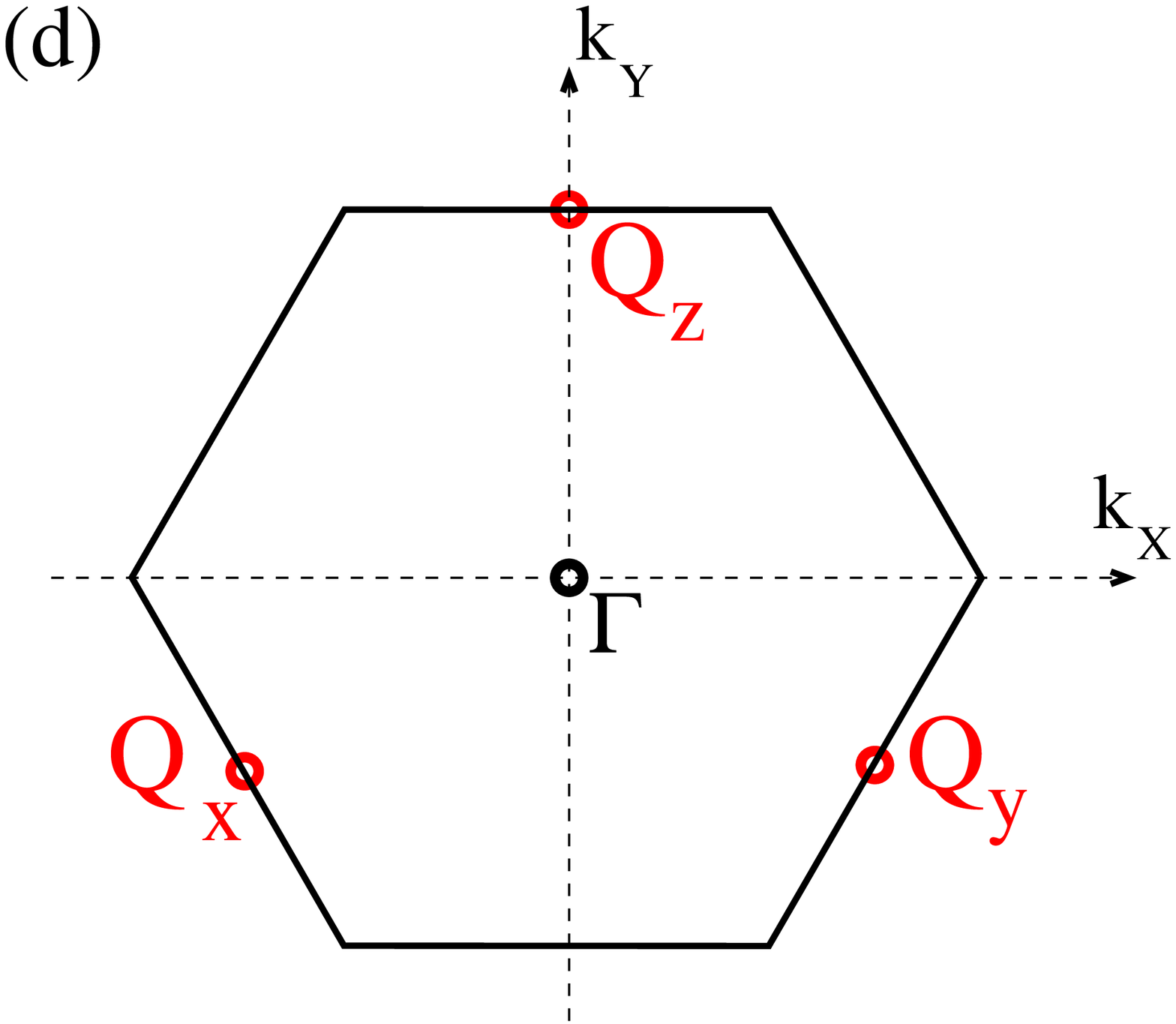}\\
\caption{\label{struc} (color online) 
(a) Schematic picture of an octahedron (6 oxygens
  surrounding a Ir$^{4+}$ ion) with the octahedral axes $x$, $y$, $z$ - which
  are easy axes on identically-labeled bonds in the Kitaev-Heisenberg model.
(b) A hexagon in the Ir$^{4+}$ plane, indicating bonds labeled $x$,
$y$, $z$ according to their local easy axis regarding Kitaev interactions.
(c) Honeycomb (direct) lattice with the labeling
  $x,y,z$ of bonds according to their in-plane orientation. A unit cell
  (centered on a $z$-bond) is represented in the dashed rectangle, as well 
  as elementary translations $\vec n_{1/2}$.
(d) First Brillouin zone, with the position of stripe-order wave vectors
  $\vec{Q}_\gamma$ ($\gamma=x,y,z$).}
\end{center}
\end{figure}

\begin{figure}
\begin{center}
\includegraphics[width=4.2cm]{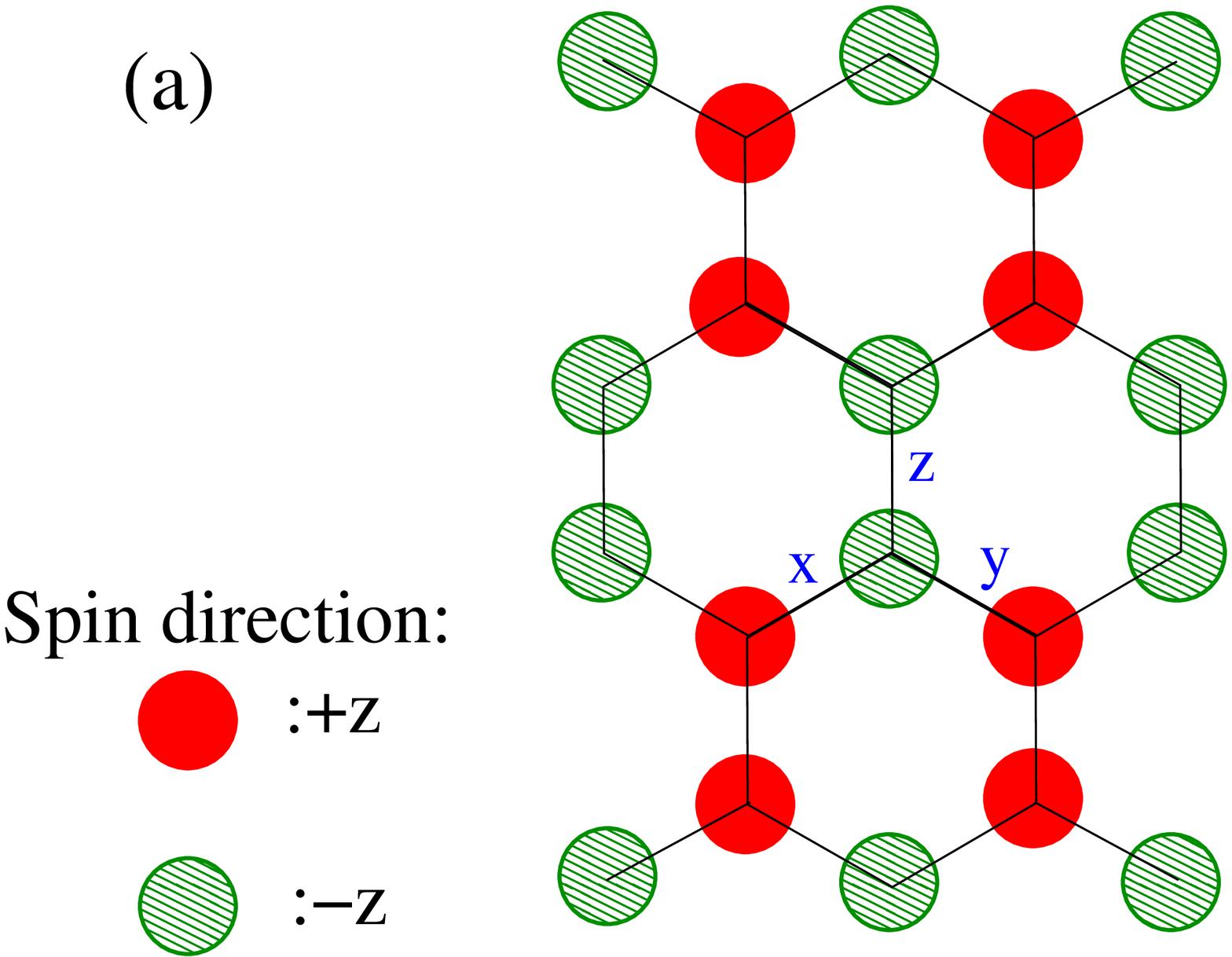}
\includegraphics[width=4.2cm]{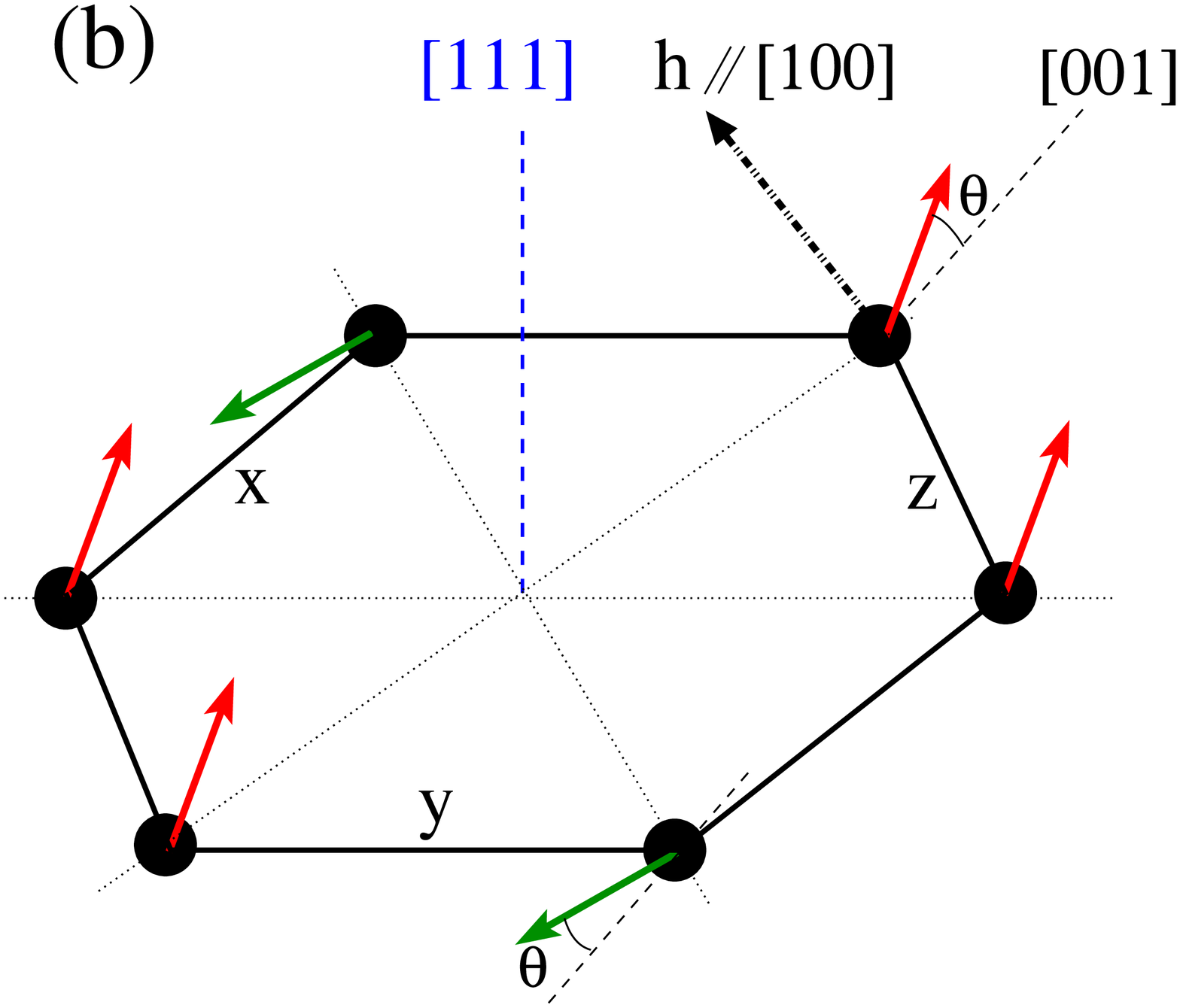}\\
\caption{\label{stripe2} (color online) (a) Representation of a $z$-stripe
  pattern (stripe pattern with spins oriented along the $z=[001]$ axis -
  neighboring spins are aligned on $z$ bonds and antialigned on other
  bonds. (b) Schematic representation, at the level of a single hexagon, 
  of a canted stripe pattern: here a $z$-stripe pattern, with spins tilted 
  of an angle $\theta$ towards a field along [100] direction.}
\end{center}
\end{figure}

We first want to characterize the various phases found in a magnetic field. It
is already known \cite{CJK} that at zero field, the spin liquid phase extends
from $\alpha_{S/L}\simeq 0.80$ to $\alpha=1$ (the Kitaev limit); the
stripe-ordered phase, with spins pointing either in $x$, $y$ or $z$ direction
[for the last case, the spatial structure, i.e. $z$-stripe pattern, is shown in 
Fig.~\ref{stripe2}(a)], is located between $\alpha_{N/S} \simeq 0.40$ and
$\alpha_{S/L}$; and a N\'eel-ordered phase is found for 
$0\le\alpha<\alpha_{N/S}$. To determine whether these phases persist in a
magnetic field and to understand the magnetic response of the system, one
needs adapted structure factors which act as order parameters. For the 
N\'eel phase, it reads:
\beq
S_{Neel}=\frac{1}{N^2}\sum_{\vec r, \vec r'} \nu_{\beta,\beta'} \langle 
\vec{\sigma}_{\vec r, \beta} \cdot \vec{\sigma}_{\vec r , \beta'} \rangle
\eeq
with $\nu_{\beta,\beta'}$ being 1 for $\beta=\beta'$, and $-1$ otherwise; 
for the stripe-ordered phase, the structure
factors corresponding to $x$- and $z$- stripe patterns are defined as follows:
\bea
S^x(\vec{Q}_x)&=&\frac{1}{N^2}\sum_{\vec r ,\vec r',\beta,\beta'}
e^{i \vec{Q}_x \cdot (\vec r' -\vec r )} \nu_{\beta,\beta'} \langle \sigma^x_{\vec r,
  \beta}
\sigma^x_{\vec r', \beta'} \rangle, \\
S^z(\vec{Q}_z)&=& \frac{1}{N^2}\sum_{\vec{r},\vec r',\beta,\beta'} 
e^{i \vec{Q}_z \cdot (\vec r' - \vec r)}
\langle \sigma^z_{\vec r, \beta} \sigma^z_{\vec r , \beta'} \rangle, 
\eea
with $\vec{Q}_{x/y/z}$ shown in Fig.~\ref{struc}(d); for instance, 
$\vec{Q}_z=(0,\frac{2\pi}{\sqrt{3}})$ which is equivalent to the more
intuitive $\vec{Q'}_z=(2\pi,0)$ (but which is not in the first Brillouin
zone). Both wave vectors account for a phase factor $-1$ between neighboring
unit cells separated by $(\pm 1/2, \pm \sqrt{3}/2)$ in real space. These
structure factors also evidence the symmetries spontaneously broken in each of
the 6-fold degenerate ground states in the stripe phase: indeed at zero field, 
in addition to translational symmetries, the Hamiltonian possesses a 
$\mathbb{Z}_2*\mathbb{Z}_3$ symmetry - $\mathbb{Z}_2$
standing for the $\vec{\sigma}_i \rightarrow -\vec{\sigma}_i$ transformation 
and $\mathbb{Z}_3$ (labeled as $C_3^*$ in Ref.~\onlinecite{JGQT}) for the 
invariance by a cyclic permutation of the $x$, $y$ and $z$ spin components
coupled to a spatial rotation by $2\pi/3$. 

Both AF (N\'eel and stripe) ordered phases of the model persist 
under moderate fields, which tilt spins by an angle proportional to the 
small field amplitude (\textit{canted N\'eel phase} \cite{ashm} or 
\textit{canted stripe phase} \cite{JGQT}). To allow for this tilting, 
the untilted spins (for $h$ infinitesimal) should not be collinear with 
the field \cite{tilt}, and the staggered magnetization of the selected ordered 
pattern has an angle with $\vec h$ as close to $\pi/2$ as possible.

We first discuss the effect of a small field in the N\'eel phase, which 
allows to precise what we will understand in the following by (an)isotropy in
the magnetic response. In the pure Heisenberg model, both classical and
quantum cases, spontaneous breaking of $SU(2)$ symmetry in the thermodynamic
limit leads to an anisotropic zero-field susceptibility. On the other hand,
the ground state of a finite cluster is known to be a singlet, which does not
break $SU(2)$ symmetry. In order to perform a classical treatment of the
Heisenberg model in absence of spontaneous symmetry breaking, one can consider
for instance a $SU(2)$ symmetric combination of classical N\'eel states - such a
state can mimic, from an experimental viewpoint, the response of a
polycristalline sample. The magnetic response of this state is isotropic in
the sense that a small field results in a magnetization of amplitude $m_0$
along the field, and $m_0$ does not depend on the field direction.
When considering the Kitaev-Heisenberg model for small finite $\alpha$, with
the classical approach above, one can find for any field direction a N\'eel 
pattern with spins perpendicular to it; within this approximation, the magnetic 
response is again isotropic, despite the absence of $SU(2)$ symmetry due to 
small Kitaev interactions. However, the combination of quantum fluctuations and 
small Kitaev interactions have been found to select one of the cubic axes as 
an easy spin axis and open thereby a small spin gap \cite{CJK}; this 
will lead to small anisotropies in the spin susceptibility. 

\begin{figure}
\begin{center}
\includegraphics[width=7.5cm]{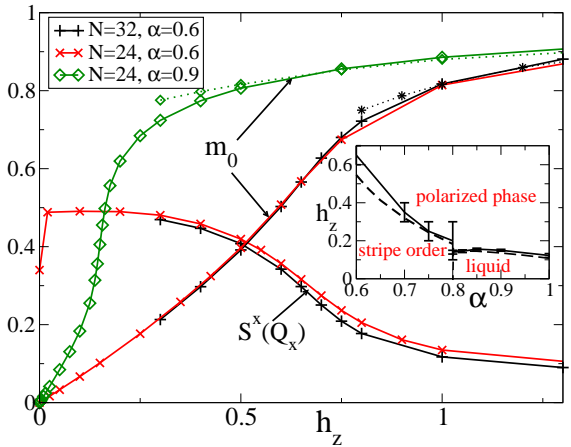}
\caption{\label{diag} (color online) Magnetization $m_0$ (per site) and 
$x-$stripe structure factor $S^x(\vec{Q}_x)$ versus $h_z$ for various values of
$\alpha$ and $N$. 
Dotted lines: Perturbative results for $m_0$ in the polarized phase.
Inset: Phase diagram in the $(\alpha,h_z)$ plane; continuous transition lines
are interpolations of points from $N=24$ data and dashed lines are
perturbative estimates.} 
\end{center}
\end{figure}

In the stripe phase, even at the classical level only six ordered patterns 
are allowed at zero field. For a generic field $\vec h$ with $h_x$, $h_y$ 
and $h_z$ taking three distinct values, the three allowed stripe orientations 
are energetically differentiated, depending on the field direction. 
For instance, in a small field $h \parallel [100]$, the two $z$-stripe patterns 
are favored (along with the $y$-stripe patterns), since they allow for the field 
to tilt spins by a small angle $\theta$ [see Fig.~\ref{stripe2}(b)], such that
the uniform magnetization develops an $x$-component. This effect 
is evidenced by structure factors characteristic of the ordered phase - in
the case of Fig.~\ref{diag} (with $\alpha=0.6$), in a small $[001]$-oriented 
field the $x$-stripe pattern becomes more favorable than the $z$-stripe 
pattern: $S^x(\vec{Q}_x)$ jumps from approximatively $1/3$ at $h=0$ to $1/2$
at small finite $h_z$, while $S^z(\vec {Q}_z)$ (not shown) drops to almost 
$0$. As for the magnetization per site $m_0(h_z)$, it grows linearly with
$h_z$ in this regime.

\begin{figure}
\begin{center}
\includegraphics[width=7.8cm]{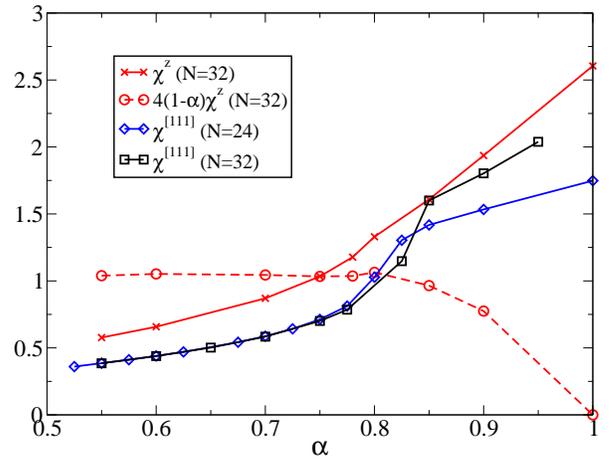}\\
\caption{\label{khiz1} (color online) 
Susceptibility per spin $\chi^z$, evaluated as
  $m_0/h_z$ for small $h_z=0.01$, and its equivalent $\chi^{[111]}$ for a
  $[111]$-oriented field, on clusters $N=24$ or $N=32$. Shown also the quantity
$4(1-\alpha)\chi^z$ for $N=32$.}
\end{center}
\end{figure}

The tilting angle $\theta$ can be determined simply in the classical approach, 
where the energy per unit cell as function of $\theta$, for given $\alpha$ and 
$h_z$, is:
\beq
E_S(h_z,\theta)=-(1+\alpha)-2h_z\sin\theta+4(1-\alpha)\sin^2\theta ,
\eeq
and its minimization w.r.t. $\theta$ gives a magnetization per
spin $m_0=\sin\theta_0=\frac{h_z}{4(1-\alpha)}$, in relatively good agreement
with the linear behavior observed at small field (see Fig.~\ref{diag}
for the case $\alpha=0.6$); also, the predicted $\alpha$-dependence of
the (uniform) susceptibility per spin,
\beq
\chi^z=\Big(\frac{d m_0}{d h_z}\Big)_{h=0}=\frac{1}{4(1-\alpha)},
\label{chz}
\eeq
is in good agreement with the numerical results in Fig.~\ref{khiz1}. 
Note that the direction of the field is relevant in the stripe
phase: a field $h\parallel[001]$ differentiates stripe
patterns, and selects two of them where spins are exactly orthogonal
to the field as $h \rightarrow 0$ ; in contrast, in a $h\parallel[111]$ field 
the three stripe directions
remain equivalent (thus $S^x(\vec{Q}_x)=S^z(\vec{Q}_z)\simeq 1/3$ for $h
\rightarrow 0$); also, none of the uncanted stripe patterns is orthogonal to a
$[111]$-field, so the magnetic response for this field direction necessarily
differs from the response to a $[001]$-field. 

\begin{figure}
\begin{center}
\includegraphics[width=7.8cm]{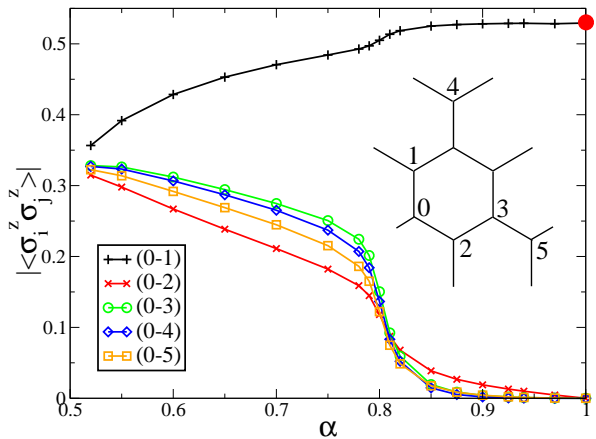}
\caption{\label{diag2} (color online)
Evolution of correlations $|\langle \sigma^z_i \sigma^z_j\rangle|$ with 
$\alpha$ on a $N=24$ cluster at $h=0$, for various relative positions $(i-j)$
with $i$ and $j$ indicated in the inset. The red dot indicates the value $\sim
0.53$ of the $(0-1)$ correlation at $\alpha=1$, which coincides with an
analytical result from Ref.~\onlinecite{Bask}.} 
\end{center}
\end{figure}

We now turn to the spin liquid regime, corresponding to the weakly perturbed
Kitaev model, and describe the effect of perturbations to the system. 
Our main focus is on the $\langle \sigma_i^z \sigma_j^z \rangle$ correlations.
In the pure Kitaev model ($\alpha=1$, $h=0$), these correlations are non-zero
only when $i,j$ are nearest neighbors on a $z$-bond \cite{Kit,Bask}, 
where they take a value 
$\langle \sigma^z_i \sigma^z_j\rangle_z^{(\alpha=1, h=0)}\simeq 0.53$ which we
can see in Fig.~\ref{diag2} (the finite-size correction to this quantity, 
i.e. to the ground state energy, is negligible even for the $N=24$ sites 
cluster).

The gapless nature of excitations persists in a small axial field
$h_z$; in contrast, a $[111]$-oriented field opens a gap resulting in a
topologically ordered phase \cite{Kit,JGQT}. Yet even in the first case,
the field causes peculiar $\langle \sigma_i^z \sigma_j^z \rangle$ 
correlations with spatial oscillations perpendicular to 
$z$-bonds and a power-law decaying envelope, as shown for $\alpha=1$ in
Ref.~\onlinecite{Tik}. Some of these trends appear in the spin-spin
correlations displayed in Fig.~\ref{cormag}(a). Due to the small size, one
cannot identify a power-law behavior; but one notices that correlations
decrease with distance, much more slowly in the 
direction parallel to $z$ bonds than in the direction perpendicular to them 
(i.e. parallel to $\vec{n}_1 - \vec {n}_2$). Given the periodic
boundary conditions on the $N=24$ lattice, this is in agreement with the
period of three unit cells for spatial oscillations predicted in
Ref.~\onlinecite{Tik}. The liquid phase is stable with respect to small
Heisenberg interactions (for $\alpha > \alpha_{S/L} \simeq 0.80$ at zero 
field \cite{CJK}). These interactions induce correlations between further
neighbors, of different type than those induced by a small field [compare 
Fig.~\ref{cormag}(a) and (b)]. The spatial alternation of sign on 
Fig.~\ref{cormag}(b) reminds of the $z$-stripe pattern, with the difference
that here these correlations are short-ranged. A proof of this last point has 
been given in Ref.~\onlinecite{MBS} for $h=0$, and indeed absolute values 
$|\langle \sigma^z \sigma^z\rangle|$ of correlations shown in Fig.~\ref{diag2} 
seem to decrease rapidly -- possibly exponentially -- with distance, at least 
in the spatial direction orthogonal to $z$ bonds. On nearest neighbors, these 
correlations are proportional to $1-\alpha$ (not too close to
$\alpha_{S/L}$). With both perturbations at play, i.e. when both $1-\alpha$
and $h_z$ are finite and small, their effects should add up; therefore, and 
since the correlations induced by Heisenberg terms decay exponentially, one 
should observe a correlation pattern with a power-law tail as in the case of 
$\alpha=1, h_z\ne 0$, but modulated as in a $z$-stripe pattern for short 
distances, at least for $h_z^2 \ll 1-\alpha$. 

Eventually, for any value of $\alpha$, a sufficiently high field drives the
system into a \textit{polarized phase} with spins parallel to
$\vec{h}$. In the case of a $[001]$-field, by considering transverse 
interactions $\sigma^x_i \sigma^x_j$ and $\sigma^y_i\sigma^y_j$ as perturbations
[the unperturbed ground state being fully polarized for $h_z>2(1-\alpha)$], 
within second order perturbation theory one finds
\beq
m_0 = \langle \sigma_z \rangle\simeq 1-\frac{\alpha^2}{(h_z+4\alpha-2)^2}\;,
\eeq
in good agreement with numerical data for $h_z \gtrsim 1$ (see dashed curves
in Fig.~\ref{diag}).

\subsection
{Transitions between liquid, stripe-ordered, and polarized phases} 

When a magnetic field $h_z$ is applied to a system which is initially stripe- or
N\'eel ordered, aside from the spin canting in the AF ordered state, the
excitation energy of a competing polarized state decreases with $h_z$ so that
this state becomes the ground state above a critical field $h_{S/F}$. This
field can be evaluated by estimating the energies (both per unit cell) 
$E_S(h_z)$ of the canted stripe phase and $E_{Fz}(h_z)$ of the polarized
phase, as a function of field - then the equation $E_S=E_{Fz}$ determines $h_{S/F}$.
For both quantities, we add to the classical value a term (last terms in both
lines of Eq.~\ref{ecantpol}) accounting for quantum fluctuations, considered in
second order perturbation theory, and evaluated respectively from the uncanted
stripe- or the fully polarized state:
\bea
E_{S}(h_z)&=&-(1+\alpha)-\frac{1}{4}\frac{h_z^2}{1-\alpha}
- \frac{(1-2\alpha)^2}{\alpha},\nn\\
E_{Fz}(h_z)&=&-2h_z+3-5\alpha -\frac{2 \alpha^2}{h_z+4\alpha-2}.
\label{ecantpol}
\eea
We compare this estimate of $h_{S/F}$ with numerical estimates obtained
from the $h_z$-dependence of the stripe structure factor $S^x(\vec{Q}_x)$
and the magnetization $m_0$ for $N=24$ and $N=32$ clusters -- both quantities 
show very weak size-dependence (see Fig.~\ref{diag}) and their derivatives 
with respect to $h_z$ are maximized for values very close to each other. 
For $\alpha=0.6$, this
numerical estimation gives $h_{S/F}=0.65(5)$ compared to a value $0.55$ found
from the criterion $E_S=E_{Fz}$. Both types of estimates are shown as
a function of $\alpha$, in the phase diagram in the inset of Fig.~\ref{diag}.
For $\alpha>\alpha_{S/L}$, the magnetization curve $m_0(h_z)$ on a finite
cluster at fixed $\alpha$ ($\alpha=0.9$ in Fig.~\ref{diag}) shows a sharp 
increase in a field range centered on $h_{L/F} \simeq 0.15(1)$; this value 
is used to locate the transition to the polarized phase (our precision on 
$h_{L/F}$ is limited here by larger finite size effects in the liquid state 
than in ordered phases). This can be understood by modeling the energy per unit
cell for the liquid phase by
\beq
E_{L}=3(1-3\alpha)\langle\sigma^\gamma \sigma^\gamma
\rangle_\gamma^{(\alpha=1,h=0)}-\chi^z h_z^2 \;,
\eeq
and looking for the field value such that $E_L=E_{Fz}$. This gives a value
$h_{LF} \simeq 0.13(3)$, which increases slightly with $1-\alpha$. Similarly, 
the position of the stripe-liquid transition, which at zero field occurs at
$\alpha_{S/L}$, is almost insensitive to a small field $h_z$.

The schematic phase diagram in Fig.~\ref{diag} has a similar structure
as its counterpart for a $[111]$-oriented field \cite{JGQT}, except that we do
not see a multicritical point at $\alpha=\alpha_{S/L}, h_z=0$. This may be
connected with the absence of $C_3^*$ symmetry in the Hamiltonian
Eq.~\ref{ham} in a $[001]$-field; moreover, the nature of the spin liquid for
$\alpha>\alpha_{S/L}$ in a small axial field differs from that of the
topologically ordered phase in a small $[111]$-field, mainly in the structure
of the low-energy spectrum.  

\subsection{Susceptibility anisotropy}

We analyze now in more detail the zero-field susceptibility: it has unusual
properties due to a special spin symmetry of the model, which is lower than
$SU(2)$. In the stripe phase ($\alpha_{N/S}<\alpha<\alpha_{S/L}$), if we
consider the general case of a $[mnp]$-oriented field, that is $\vec h$
parallel to $m \vec x + n \vec y + p \vec z$, the corresponding susceptibility
$\chi^{[mnp]}$ behaves as $\frac{c_{mnp}}{1-\alpha}$, where
$c_{mnp}$ is a rational number which can be found from the classical
picture of a spin-flop mechanism. One can see in Fig.~\ref{khiz1} that
susceptibilities along different axes follow well this
type of law in the stripe phase.
The coefficient $c_{mnp}$ is maximized for a $[001]$-field ($c_{001}=1/4$,
see Eq.~\ref{chz}), but also for a $[10\bar{1}]$-field, and minimized for a
$[111]$-field ($c_{111}=1/6$). Note here that the $[111]$ axis corresponds, in
the context of susceptibility measurements in Ref.~\onlinecite{SG}, to the $c$
axis of Na$_2$IrO$_3$, i.e. perpendicular
to the Ir-honeycomb layers [see Fig.~\ref{struc}(b)]. Concerning the
susceptibility in the $ab$ plane, this model predicts that -- in the stripe
phase -- $\chi$ takes a value between $\chi^{[001]}=\frac{1}{4(1-\alpha)}$ and
$\chi^{[11\bar{2}]}=\frac{1}{5(1-\alpha)}$. An average over the field
orientations in the $ab$ plane gives 
$4(1-\alpha)\langle\chi^{ab}\rangle=2/3+\frac{\sqrt{3}}{2\pi} \simeq 0.94$.

No such simple expressions for the susceptibility are available for the liquid
phase; but there, despite large finite-size effects (compared to the stripe 
phase) one notices that for a given axis susceptibilities tend to increase 
with $\alpha$, with again $\chi^{[111]} \le \chi^{[001]}$. As for the N\'eel 
phase, a classical approach does not predict there any anisotropy in the 
static susceptibility (see related discussion in Section \ref{phas}), and the
anisotropy resulting from quantum fluctuations at small finite $\alpha$ should 
be smaller than in the stripe phase -- but the N\'eel phase is obtained for 
values of $\alpha$ likely unrealistic for Na$_2$IrO$_3$ \cite{CJK} and also 
not supported by experiment\cite{Liu11}. 

The experimental data of Ref.~\onlinecite{SG} reveal an anisotropy
in the static susceptibility as well, but the measured susceptibility is
significantly larger along the $c$ axis than in the $ab$
plane, in contrast to our model prediction $\chi^c<\chi^{ab}$. 
This suggests that other factors could contribute to the susceptibility
anisotropy\cite{VVC}. For instance, 
anisotropic contributions are expected to arise from the Van Vleck
susceptibility $\chi_{VV}^{(ab)}\ne\chi_{VV}^{(c)}$, and also via the g-value
anisotropy in the ground-state Kramers doublet, i.e., $g_{ab}\ne g_{c}$, 
due to lattice distortions lowering the octahedral symmetry.
These effects involve the inclusion of higher, excited spin-orbital, states.

Nevertheless, we believe that the Kitaev-Heisenberg model describes some
characteristic features of layered iridates: for instance, recent
susceptibility measurements\cite{SMG} on polycrystals of Li$_2$IrO$_3$ and of
Na$_2$IrO$_3$ showed that the zero-temperature susceptibility was about twice
higher for the first (Li) compound than for the second. Assuming that both
compounds can be described by the Kitaev-Heisenberg model and are in the 
stripe phase, the value of $\alpha$ for the Li compound should be 
significantly larger than for the Na compound - probably 
$\alpha_{Li} \geq 0.6$ if one compares the paramagnetic Curie temperatures 
of both compounds and follows theoretical results of Ref.~\onlinecite{RTT}. 
This is actually in agreement with the law $\chi \propto \frac{1}{1-\alpha}$ 
which we find for any field direction (and thus for polycrystalline samples): 
since by assumption $\alpha_{Na} \geq \alpha_{N/S} \simeq 0.4$, the observed 
ratio\cite{SMG} $\chi(Li)/\chi(Na) \simeq 2$ would imply $\alpha_{Li} 
\geq 0.7$. This suggests that the Li-compound is quite close to the 
stripe-liquid transition.

In the following, we shall confine our study to the Kitaev-Heisenberg model and
explore the role of non-magnetic vacancies, unavoidably present 
in real compounds, and which, in general, can significantly alter 
the properties of frustrated magnetic systems \cite{SL2,DMNM}.

\section{\label{ces}Combined effect of a single vacancy and 
a magnetic field on spin correlations}

We now study the effect of a single non-magnetic vacancy on the magnetic
response of the system, especially at low-field. As previously, we will
address separately the different phases of the model.
\begin{figure}
\begin{center}
\includegraphics[width=4.cm]{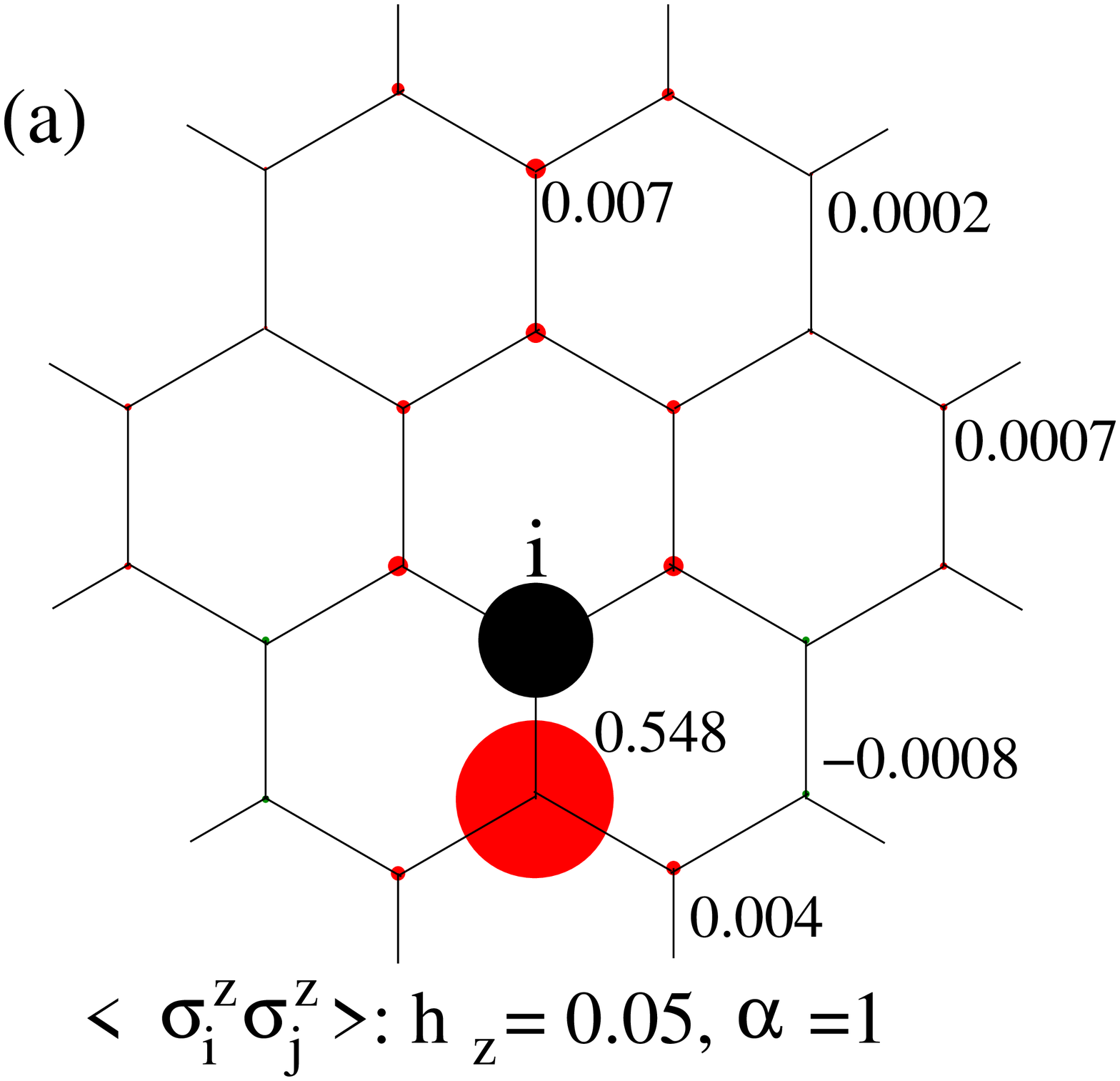}
\includegraphics[width=3.85cm]{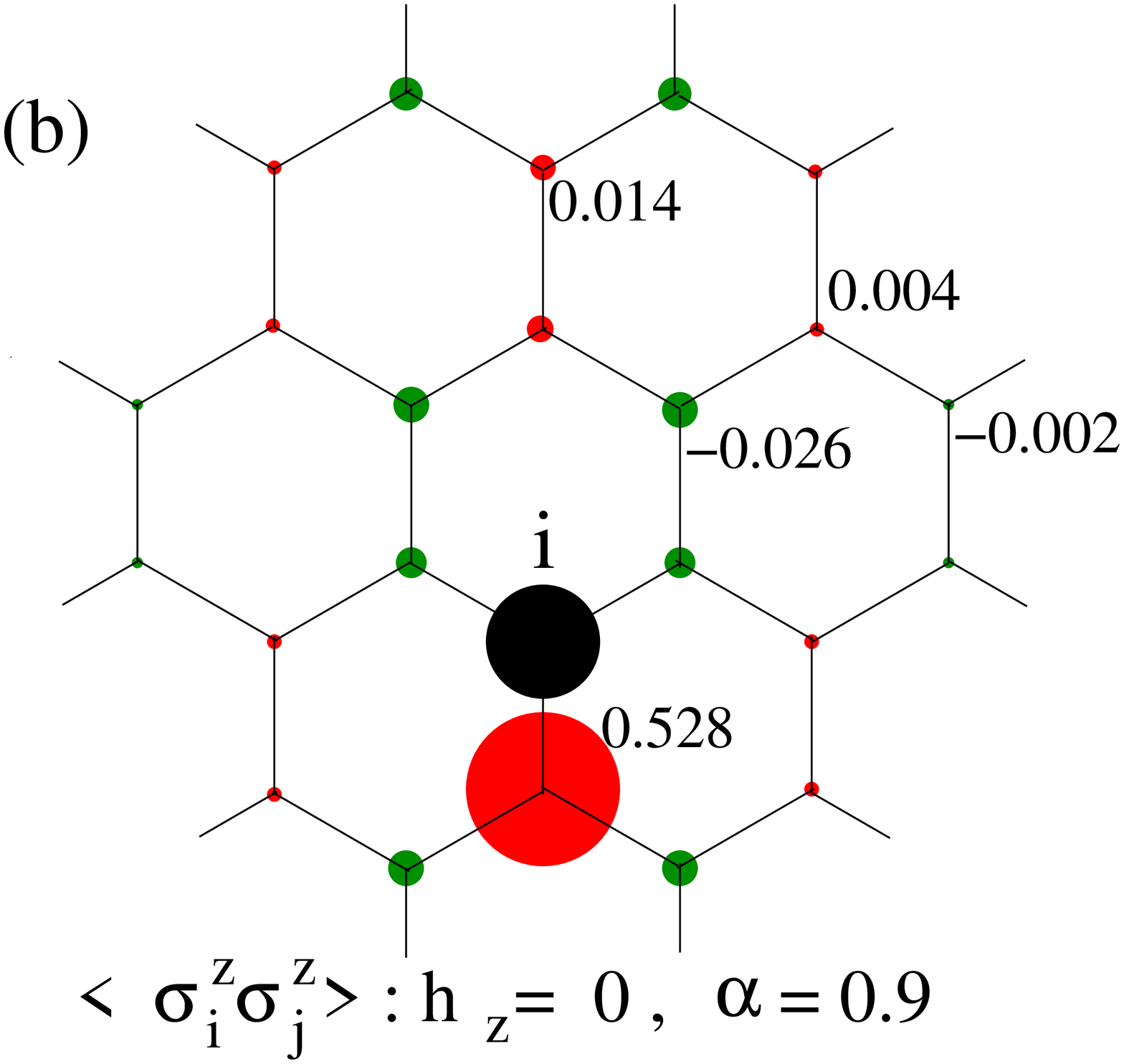}\\
\vspace{0.2cm}
\includegraphics[width=3.65cm]{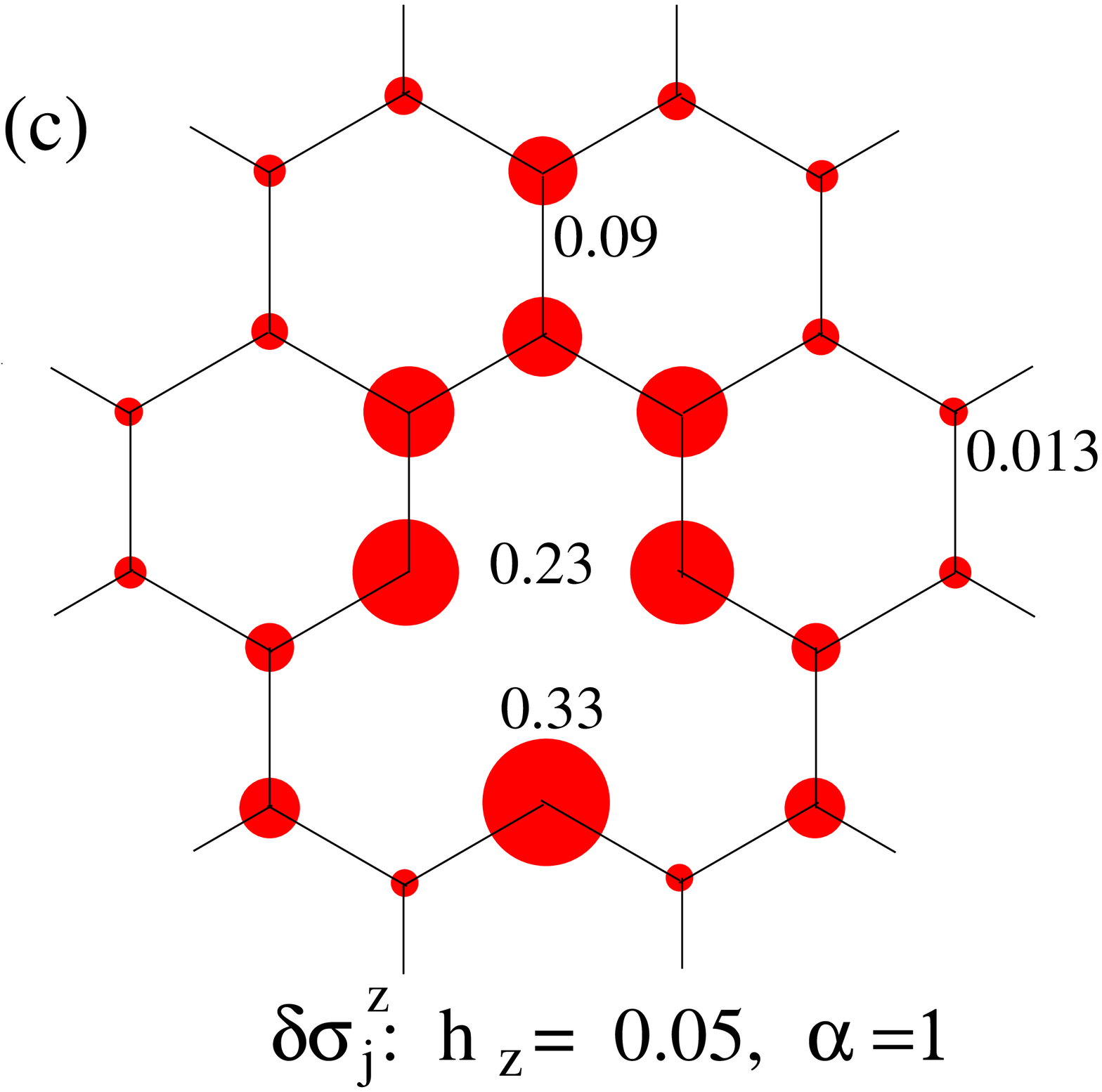}
\hspace{0.0cm}
\includegraphics[width=3.75cm]{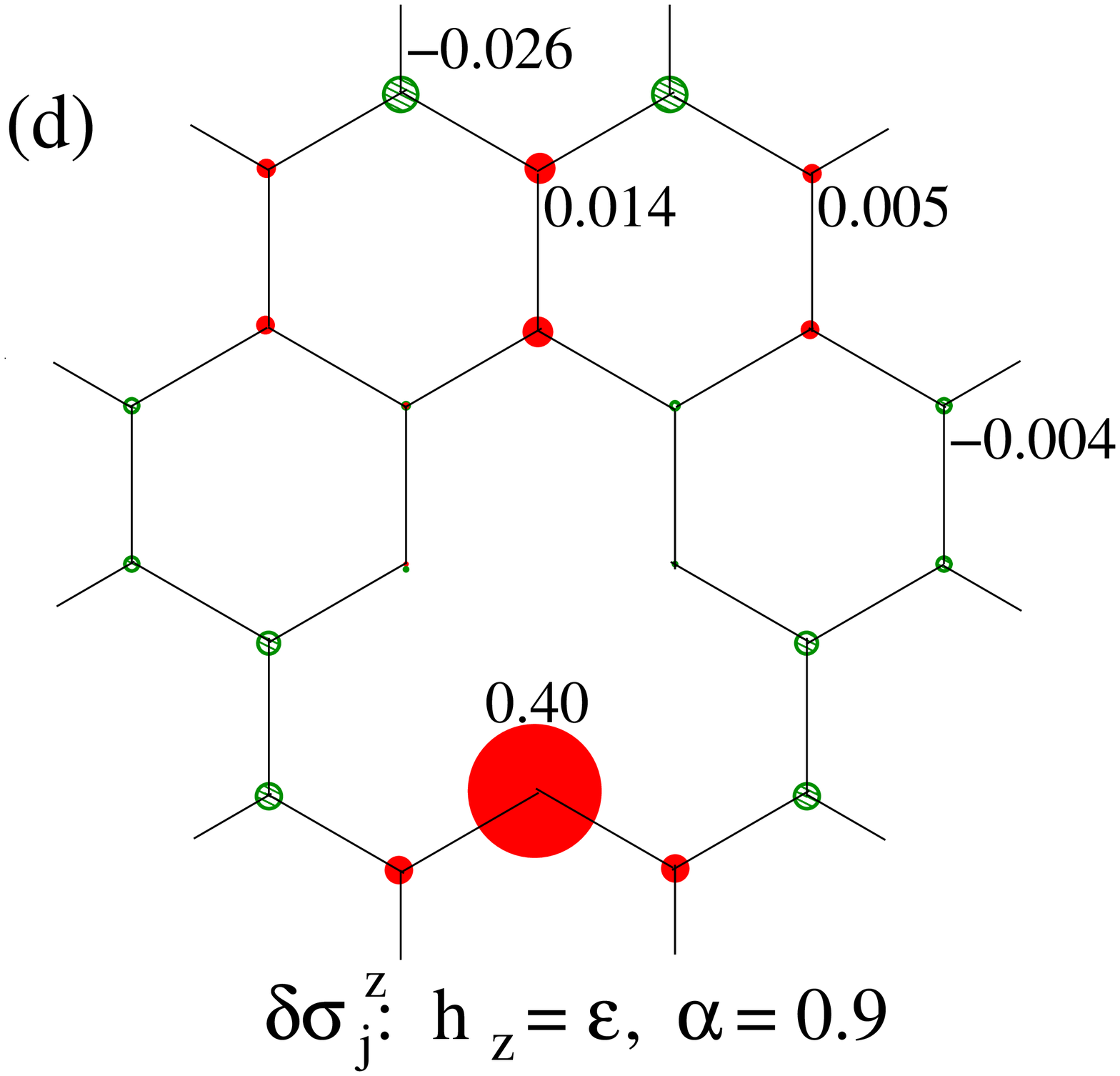}\\
\caption{\label{cormag} (color online) (a, b) Correlation functions 
$\langle \sigma_i^z \sigma_j^z \rangle$  with respect to site $i$ indicated by
  a black circle. (c, d) Induced magnetization patterns 
$\delta \sigma^z_j $ due to the presence of a
  vacancy, on the $N=24$ cluster.
  Circle areas are proportional to the quantities shown. Parameters are
 respectively $(\alpha,h_z)=(1,0.05)$ in (a) and (c), while 
$(\alpha,h_z)=(0.9,0)$ in (b) and (d) [a small symmetry-breaking 
field $h_z=\epsilon=5.10^{-4}$ was added in (d)].}
\end{center}
\end{figure}

\subsection
{Anomalous magnetization around a vacancy in the liquid phase}

\begin{figure}
\begin{center}
\includegraphics[width=7.8cm]{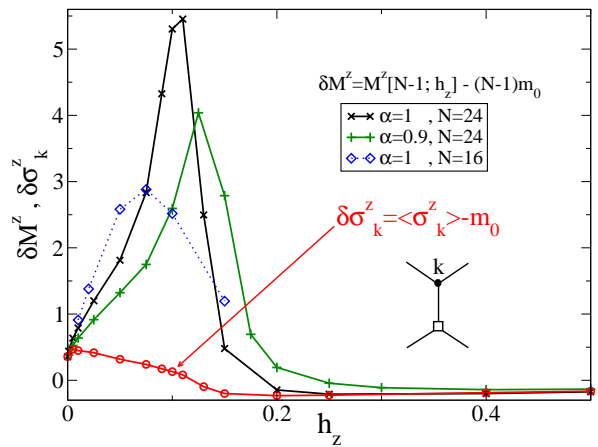}
\caption{\label{difmz} Spin-vacancy induced magnetization $\delta M^z$ as a
  function of magnetic field $h_z$, for different $\alpha$ and cluster
  sizes. Shown also is a partial contribution (open circles) of the spin at
  site $k$, being the $z$-neighbor of the vacancy, for $\alpha=1$ and $N=24$.}
\end{center}
\end{figure}

We start by considering the liquid phase, and choose the field direction
along the $[001]$ axis, for which features characteristic of the 
Kitaev model are the clearest.
First, we show that the combination of an infinitesimal field and a spin
vacancy induces a peculiar magnetization pattern around the vacancy. The
relevant quantity is the induced magnetization:
\beq
\delta \sigma^z_j = \langle \sigma^z_j \rangle -m_0 ,\nn
\eeq
which gives, for a given field $h_z$ and site $j$, the change of the 
magnetization at this site relative to its value ($m_0$) in the vacancy-free
system. We relate the vacancy-induced magnetization pattern to the correlation 
function $\langle \sigma_i^z\sigma_j^z\rangle$ of the vacancy-free system. 
The most striking feature of the magnetization patterns shown
in Fig.~\ref{cormag}(c,d) is the large magnetization on the $z$-neighbor of the
vacancy (referred to as site $k$ in Fig.~\ref{difmz} from now on). The
magnetization is significantly larger there than on any other site -- and 
only on this site does
it remain finite at vanishing field in the Kitaev model:
$\langle \sigma^z_j \rangle \rightarrow \delta_{j,k} m_N$ for $\alpha=1$ and
$h_z\rightarrow 0$ ($m_N\simeq 0.48$ for $N=24$).
Actually, in the thermodynamic limit (TL) the magnetization (both the total
one and $\langle \sigma^z_k \rangle$) should behave with field as\cite{Wil} $
h_z \ln(h_z)$; our results are consistent with this prediction in the sense that
$m_N$ decreases with increasing $N$, so that it should vanish in the TL as
expected. As for the logarithmic field-dependence, we stress that this feature
pertains to the TL, while for finite systems we can only notice weak
non-linearities in the magnetization curves, discussed later in more detail.

The magnetic response localized on the $z$-neighbor of the vacancy is
clearly connected with the fact that the correlation function
$\langle \sigma_i^z \sigma_j^z\rangle$ is non-zero only between 
nearest-neighbors sharing a $z$-bond \cite{Kit,Bask}: if one spin
is removed at site $v$, the term $\sigma^z_v\sigma^z_k$ obviously disappears
from the Hamiltonian and thus the spin $k$ is easily polarizable by the
$z$-oriented field (one can also interpret this behavior in terms of dangling
Majorana fermions \cite{Wil}). 

Yet, we already saw in Section \ref{phas} that a small field induces
$\langle \sigma_i^z \sigma_j^z\rangle$ correlations beyond nearest neighbors,
it makes sense to focus in more detail on the magnetization pattern around a
vacancy at small finite $h_z$. Fig.~\ref{cormag}(c) shows the variation of 
magnetization $\delta\sigma^z_j$ due to the 
presence of the vacancy. Indeed, the spatial variations of this quantity 
presents similarities with those of correlations in the vacancy-free 
system: $\delta\sigma^z_j$ is positive
everywhere and decreases with increasing distance to the vacancy. Moreover,
the spatial oscillations of correlation function appear in the magnetization
pattern. Indeed, if one compares values of $\delta \sigma^z_j$ at relative 
position either $\vec{n}_1+\vec{n}_2$ or $2\vec{n}_1-\vec{n}_2$ from the 
vacancy [see values $0.09$ and $0.013$ respectively in Fig.~\ref{cormag}(c)], 
their respective ratio is comparable to that of corresponding correlations 
in the vacancy-free system.

While $\langle \sigma^z_j \sigma^z_i\rangle$ correlations
scale as $h_z^2$, the magnetization enhancement $\delta \sigma^z_j$ on
further neighbors of the vacancy grows approximately linearly with
$h_z$; in contrast, for $j=k$, it decreases from
$m_N$ at $h=0$ towards small negative values in the polarized phase (see
$\delta \sigma^z_k$ in Fig.~\ref{difmz}).
Although the response observed here is obtained for a finite cluster
size, we may expect that in a large system, the spin at site $k$ acquires a
large magnetization due to the combined effect of the neighboring vacancy and
the small field, and in turn polarizes its other neighbors in a way that
the $\delta\sigma^z_j$ reproduces the above discussed correlation
function pattern (Fig.~\ref{cormag}(a)).\\

If one considers the effect of a vacancy on the magnetic response of the
system in a field $h_z$, a relevant quantity to consider is $\delta
M^z(h_z)=M^z[N-1,h_z]-(N-1)m_0(h_z)$, with $m_0$ the magnetization {\it per
site} of the vacancy-free system, and $M^z[N-1,h_z]$ is the total magnetization
of the system with $N-1$ spins plus a vacant site. 

This quantity, shown in Fig.~\ref{difmz}, vanishes in the infinite field 
limit (where all spins are polarized), but also at small field in the TL. 
In finite systems, the previously discussed finiteness of $m_N$ goes along
with a deviation from the expected logarithmic behavior of $M^z$; yet, we
notice on large enough systems ($N=24$) a non-linearity of the magnetization
curve in the small field range $h_z\ll h_{L/F}$, the slope $\frac{d(\delta
  M^z)}{dh_z}$ decreasing moderately with increasing small $h_z$.
But when $h_z$ becomes close to the value $h_{L/F}$ of the liquid/polarized
transition for the undoped system, the slope $\frac{d (\delta M^z)}{d h_z}$
increases strongly, so that $\delta M^z$ peaks near
$h_z \simeq 0.12(2)$. The exact position of the peak depends slightly on the
cluster size, and the peak amplitude grows with increasing system size
(roughly linearly: compare amplitudes for $N=16$ and $N=24$ on Fig.~\ref{difmz}).
The existence of this peak can be understood in the following way: not only does the presence
of a vacancy cause an increase of the magnetization in the liquid phase, but
it shifts slightly the liquid/polarized transition towards smaller fields
$h'_{L/F}<h_{L/F}$ (the difference between both fields being likely
proportional to the vacancy concentration in a diluted limit), and the peak
signals a field range $h'_{L/F}<h_z<h_{L/F}$. 
This behavior can loosely remind of spin polarons formed around mobile holes 
in a paramagnetic background, 
so that the temperature of transition to the ferromagnetic (FM) phase shifts
up with increasing hole density\cite{VHJM}.
Here, instead of temperature the parameter driving the transition is the
magnetic field,
and moreover a vacancy is here not mobile, but its effect is as
well to polarize its surroundings more and more when approaching the FM phase;
consequently, at small finite vacancy concentration, the transition field
should be slightly smaller than in the vacancy-free case.

An analogy between the spin correlations in the vacancy-free system and the
magnetization pattern around a vacancy can also be noticed in the case where the
Kitaev Hamiltonian is perturbed by small Heisenberg interactions
[see Fig.~\ref{cormag}(b) and Fig.~\ref{cormag}(d)]. Note that in
Fig.~\ref{cormag}(d), we also included an infinitesimal field $h_z$, the effect
of which is only to break the Kramers degeneracy related to the odd number of
spins. There, both the correlations and the vacancy-enhanced magnetizations
$\delta \sigma^z_j$ decrease rapidly with distance to the vacancy. Moreover
the sign of these quantities alternates consistently with the pattern of
$z$-oriented stripes shown in inset of Fig.~\ref{slrstripes}, and their
decrease is much faster when going perpendicularly to stripes than along 
stripes: for the magnetizations $\langle\sigma^z_j\rangle$
(or equivalently $\delta \sigma^z_j$, since $m_0 \simeq 0$), one can again
compare in Fig.~\ref{cormag}(d) the absolute values at positions
$2\vec{n}_1-\vec{n}_2$ ($0.004$) and $\vec{n}_1+\vec{n}_2$ ($0.014$) from the
vacancy. 

The staggered $z$-bond magnetization, which measures the $z$-stripe order
parameter, 
\bea
  m^z(\vec{Q}_z)=\frac{1}{N}\sum_{\vec r} e^{i \vec{Q}_z \cdot \vec r}
  \langle \sigma^z_{\vec r,A}+\sigma^z_{\vec r,B} \rangle ,
\eea
is shown in Fig.~\ref{slrstripes}. It also evidences that, in the liquid
phase, the short-range stripe pattern is induced around the vacancy by small
Heisenberg interactions. Starting from the $\alpha=1$ limit, the
magnetizations at further neighbors increase (in absolute values) 
significantly with increasing amplitude of Heisenberg interactions, 
i.e. with decreasing $\alpha$. We expect that the extensive quantity
$Nm^z(\vec{Q}_z)$ diverges in the TL for $\alpha \rightarrow \alpha_{S/L}$.
Moreover, we note that the vacancy-induced magnetization $\delta M^z$, for
fixed small field $h_z$, is reduced by the presence of Heisenberg
interactions. It is significantly smaller for $\alpha=0.9$ than for
$\alpha=1$, as the tendency to stripe order around the vacancy, at small 
field, competes with the aligning effect of the field. Consequently, the peak
in $\delta M^z(h_z)$ shifts slightly towards higher fields with decreasing
$\alpha$: the effect of Heisenberg interactions is to increase the position
$h'_{L/F}$ of transition to the polarized phase (as for $h_{L/F}$ in the
vacancy-free case, see Fig.~\ref{diag}).
 
We briefly comment here on the fact that in the polarized phase (and unlike in
small field), the vacancy-induced magnetization change $\delta \sigma^z_k$ at
the $z$-neighbor is negative.
This can be understood with the perturbative treatment of transverse
interactions described in Section \ref{phas}; but here, to evaluate
the quantum correction to the magnetization at a given site, we consider the
off-diagonal (perturbing) couplings only on neighboring bonds -- see 
Ref.~\onlinecite{SBEFP}
for a similar approach. Within this approach, the magnetization at this site is
$\langle\sigma^z_k\rangle \simeq 1-\frac{\alpha^2}{(h_z+2\alpha-1)^2}$; and
the predicted value for $\delta \sigma^z_k$ is of the same order as seen in
Fig.~\ref{difmz} and it decreases in a similar way as $h_z$ increases,
sufficiently far away from the transition field to either the liquid or the
stripe phase. Qualitatively, one can interpret this by the fact that at high
field, fluctuations from a fully-polarized state are effectively stronger at
the vacancy's $z$-neighbor than elsewhere, since the absence of ferromagnetic
interaction on the missing $z$-bond makes it easier to flip this spin than
other spins. With a similar reasoning one can explain that, in the same field
range, the magnetization at other nearest neighbors of the impurity 
is enhanced compared to $m_0$.

If, instead of a field along an easy axis, a field in the $[111]$ direction 
is applied, one can assume\cite{Wil} that the magnetic response is 
approximately a linear combination of responses to each field component: 
thus each neighbor of a vacancy acquires an anomalously large magnetization 
- the neighbor spin at site $j$, that would be connected to the missing spin 
by a $\gamma$-bond, has a magnetization 
$\langle \sigma_j^\gamma \rangle \sim h_\gamma \ln(h_\gamma)$, and in turn 
brings a smaller polarization, also parallel to $h_\gamma$, to other spins 
in its vicinity - mainly on neighboring sites which are close to the spatial 
axis in the prolongation of the missing $\gamma$-bond.
Hence, three domains should coexist in the system, each with a different 
magnetization direction. Note that a vector 
$\vec{m}_{stripe}=[m^x(\vec{Q}_x),m^y(\vec{Q}_y),m^z(\vec{Q}_z)]$
can be defined and used to characterize the polarization of spins around the
vacancy. Qualitatively, $\vec{m}_{stripe}$ should point in a direction closer 
to the easy axis $\gamma$ for which $h_\gamma$ is the largest than to other 
easy axes; but it can have three non-zero components simultaneously.

\subsection{\label{sosf} 
Stripe orientation in a small field and anisotropy of magnetic response}
 
\begin{figure}
\begin{center}
\includegraphics[width=8.0cm]{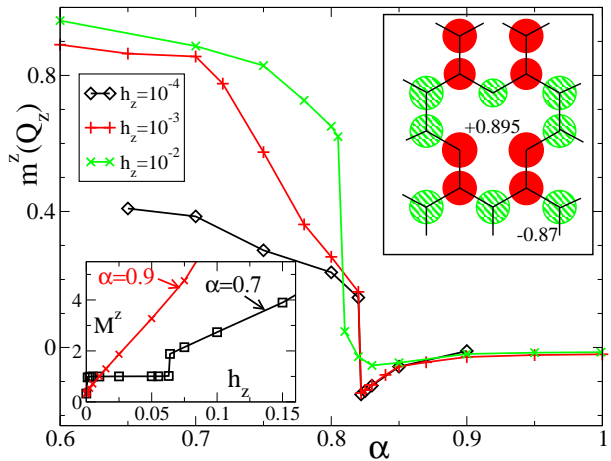}\\
\caption{\label{slrstripes} (color online) The $z$-stripe  order parameter
  $m^z({\vec {Q}_z})$ computed on the $N=24$ cluster, as a
function of $\alpha$ at different small fields $h_z$. Right-inset: The
magnetization pattern $\langle \sigma^z_{\vec r}\rangle$ 
around the vacancy for $\alpha=0.7$ and $h_z=10^{-3}$. Left-inset: Total
magnetization $M^z$ as a function of a field $h_z$ applied in the
liquid ($\alpha=0.9$) and in the stripe ($\alpha=0.7$) phases, with a
magnetization jump at $h_{an} \simeq 0.062$.}
\end{center}
\end{figure}

In the stripe-ordered phase, among the 6 allowed stripe patterns (3 possible 
stripe directions times 2 spin orientations), a vacancy in combination with 
an infinitesimal field $h_z>0$ selects the $z$-stripe pattern for which
$M^z=1$ up to small quantum corrections. This selection is due to the
necessary imbalance between the number of spins respectively aligned
and antialigned with the field (see inset of Fig.~\ref{slrstripes}).
This state is stable versus increasing $h_z$, until the field reaches
a critical value (or anisotropy field $h_{an}$), where canted stripe
patterns -- those with stripe directions along $x$ or $y$ -- become more
favorable. This effect is similar to the local rotation of the staggered 
magnetization close to a vacancy in $SU(2)$ antiferromagnets \cite{egg1}. 
Here, the main difference is that there is only a finite number of stripe 
patterns which cannot
be continuously connected to each other without going out of the ground state
manifold; consequently, the system has to choose between allowed stripe
patterns, tilted or not by the field, and $\vec{m}_{stripe}$ is forced to
point exactly along one easy axis $\gamma$, corresponding to (one of) the most
favorable pattern(s). To understand the existence of a finite anisotropy
field, one has to compare the energy of the $z$-stripe pattern
(\textit{untiltable}\cite{tilt} in a $[001]$-field) to that of the
\textit{tiltable} $x$- and $y$-stripe patterns. The former, thanks
to its total magnetization  $M^z\simeq +1$, acquires locally an energy gain 
$\simeq - h_z$ with respect to the zero-field value but it has zero
susceptibility ($\chi^z=0$) thus no contribution $\propto Nh_z^2$ to its
energy; in contrary, the field lowers the energy of the latter patterns by 
an extensive quantity $\delta E \simeq -(N-1)\frac{h_z^2}{8(1-\alpha)}$. This
comparison gives $h_{an}=8\frac{1-\alpha}{N-1} \simeq 0.10$ for $\alpha=0.7$,
not far from the position of the jump in $M^z(h_z)$ in Fig.~\ref{slrstripes}.  

In the vicinity of the stripe-liquid transition, the magnetization pattern
induced by a small $[001]$-field ($|h_z| \ll 0.1$) around a vacancy can be of
special interest: assuming that vacancies are diluted enough such that their
mutual interactions can be neglected, the patterns expected on both sides of the
transition are clearly distinct from each other. In the stripe phase, all
neighboring spins of the vacancy have a large, almost field-independent 
magnetization, with two of them aligned with the field and the third one
antialigned with $\vec h$. In contrast, in the liquid phase only one nearest 
neighbor spin is significantly polarized (with a strongly field-dependent 
polarization) and aligned with the field; and the two others are weakly
polarized opposite to the field. Thus, if one would have a local probe to
determine the (site-resolved) magnetization in the neighborhood of a vacancy
-- this probe could be $^{23}Na$ NMR for instance -- the measured pattern
could indicate on which side of the transition the system is. 

We have also seen that a vacancy can block the spin canting mechanism in the
stripe phase, under a small field in (or close to) the $[001]$ axis. This
phenomenon is absent for a field in the $[111]$ direction, where all stripe
patterns remain equivalent and continuously tiltable by the field.
Consequently, in a finite system, the situation concerning the susceptibility
anisotropy at zero-field: $\chi^z=0<\chi^{[111]}$, is now opposite to the
previously discussed case without vacancies where we found
$\chi^{[111]}<\chi^z$. Concerning now the thermodynamic limit with a finite 
concentration $n_v$ of vacancies, if one assumes that $n_v$ is
large enough to influence the magnetic response at low temperatures but small
enough so that vacancies behave independently from each other, a similar
effect could account for the experimentally observed $\chi^{ab}<\chi^c$.
However, given that in a small field, within the Kitaev model, the
magnetization enhancement caused by two vacancies residing on the same
sublattice is much larger than twice the magnetization enhancement caused by a 
single vacancy \cite{Wil}, one should take care about cooperative effects
between the vacancies.

\section{\label{ibv}Interaction between vacancies: broken 
orientational symmetry and magnetization properties}

We turn now to the case where several vacancies are present in the system; 
we first consider their effect on the magnetization curves; then, we will 
focus on their effective interaction in the stripe phase and show how a 
vacancy pair can select an ordered pattern.

\subsection{Comparison of magnetic response with vacancy pairs 
in various phases}

\begin{figure}
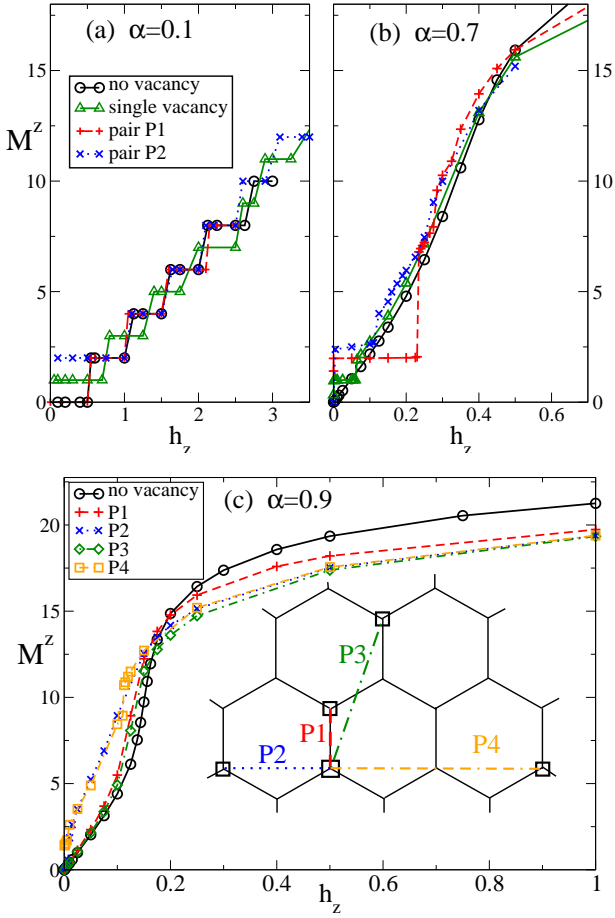

\begin{center}
\includegraphics[width=8.1cm]{compmz_a17.eps}\\
\vspace{0.24cm}
\includegraphics[width=7.8cm]{mzh2_a09N24.eps}\\
\caption{\label{magNSL} (color online) Magnetization curves $M^z(h_z)$ on the
  cluster $N=24$ with 0, 1, or 2 vacancies. The labels $P1,...,P4$ 
correspond to relative positions of 2 vacancies as shown in panel (c).
Panels (a),(b), and (c) correspond to $\alpha=0.1, 0.7, 0.9$, respectively.}
\end{center}
\end{figure}

It is instructive to compare the effect of vacancies or pairs of vacancies on
magnetization curves, for given field direction and system size. To do this 
comparison we choose a $[001]$-field and consider the periodic 
$N=24$ cluster, but discuss as well implications for the thermodynamic limit.
Magnetization curves $M^z(h_z)$ without or with vacancies, in different phases
of the Kitaev-Heisenberg model, are shown in Fig.~\ref{magNSL}.

\textit{In the N\'eel phase}, although the total magnetization
$M^z=\sum_i \langle \sigma^z_i \rangle$ is not a conserved quantity in 
presence of small
Kitaev interactions ($\alpha \ll \alpha_{N/S}$), the magnetization curve
$M^z(h_z)$ on finite clusters displays a succession of plateaux (almost flat
for $\alpha=0.1$), as seen in Fig.~\ref{magNSL}(a). They are separated by jumps 
$\Delta (M^z) \simeq +2$ corresponding to a flip of 
one spin $1/2$. At fixed $\alpha$ the step-like increase of the magnetization
per site $m_0(h_z)$ disappears only in the TL, the steps becoming smaller 
and closer to each other as $N$ increases. Moreover the steps disappear, 
at fixed cluster size, as $\alpha$ gets closer to $\alpha_{N/S}$.

Within this phase, one is tempted to define an effective spin $1/2$,
$\vec{S}_{eff}$, associated to the vacancy, such that 
$M^z(h_z\rightarrow 0)= +1$ corresponds to $\vec{S}_{eff}$ pointing along 
$+z$; with two vacancies, their effective $S_{eff}^z$'s add up if vacancies 
are on the same sublattice (e.g., n.n.n. case) and cancel each other if on 
opposite sublattices (n.n. case). From the behavior of $M^x(h_x \rightarrow
0)$ one obtains similar conclusions concerning effective couplings between 
other components of $\vec S_{eff}$. One could therefore think that an
effective interaction between two vacancy spins, if it can be defined, should 
be $SU(2)$-invariant, and possibly of Heisenberg type $\vec{S}^i_{eff}\cdot 
\vec{S}^j_{eff}$; yet, the gapless nature of the undoped system and the 
continuous behaviour of magnetization curves in the TL make it unclear how 
to formulate the effect of vacancies in terms of effective spins.

\textit{In the liquid phase}, in contrast, the magnetization does not show any plateau,
but vacancies cause peculiar features at small fields:
We have already seen that, compared to the zero-vacancy curve, the slope of 
$M^z(h_z)$ at small field is significantly increased by the presence of one
vacancy in the system. If two vacancies are present, their effect
depends strongly on their respective sublattices (see Fig.~\ref{magNSL}(c)):
if they are on opposite sublattices the $M^z(h_z)$ curve is, as in the
vacancy-free case, linear for small $h_z$ [and $\chi^z(h_z\rightarrow 0)$ is
only slightly modified]; but if they are on the same sublattice, this slope
is greatly increased -- for a large range of $h_z$ the vacancy-induced increase
in $M^z$ is more than twice that of the single-impurity case. Again, even though
we cannot identify a logarithmic behavior $\chi^z(h_z)\sim
1/(h_z\ln^{3/2}h_z)$ as derived in Ref.~\onlinecite{Wil} for
$\alpha=1$, we see that vacancies, as soon as they are not equally distributed
on both sublattices, act like \textit{partially-unbound moments} causing
non-linearities in the magnetization properties at low field.

\textit{In the stripe phase}, as in the liquid phase, Kitaev interactions 
are strong enough to prohibit the existence of plateaux in the magnetization
curves $M^z(h_z)$, except at small fields in presence of vacancies: with one
vacancy, the previously discussed $M^z=1$ plateau is again suggestive of an
effective spin $1/2$ of the vacancy; and with two vacancies on the same
sublattice, the $M^z \simeq 2$ plateau seen on Fig.~\ref{magNSL}(b) (pair P1)
extends up to a field $h_{an2} \leq 0.22(1)$, that is more than twice the
anisotropy field $h_{an}$ for a single vacancy. This indicates a cooperative
effect of both vacancies, which stabilize the $z$-stripe pattern more than if
they behaved independently from each other. To describe this cooperative
behavior, we will, similarly as in Section \ref{sosf},
analyze the effect of vacancies on the energies of the various stripe patterns.

\begin{figure}
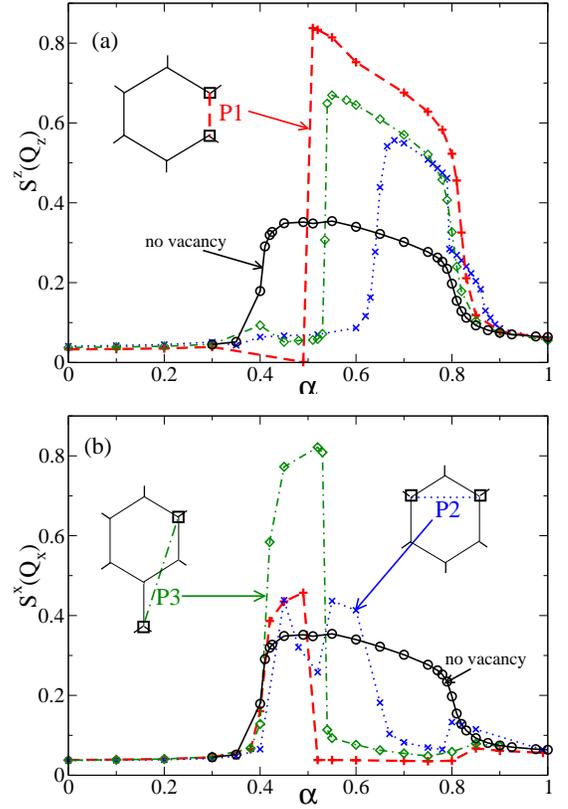

\begin{center}
\includegraphics[width=7.2cm]{szs_N24v3.eps}\\
\vspace{0.2cm}
\includegraphics[width=7.2cm]{sxs_N24v3.eps}
\caption{\label{mstripy_2} (color online) Structure factors 
(a) $S^z(\vec{Q}_z)$ and (b) $S^x(\vec{Q}_x)$ at $h = 0$ calculated for 
various respective positions of 2 vacancies $P1$, $P2$, $P3$, indicated
in the insets. Also shown is the structure factor of the vacancy-free 
system (open circles), for which $S^z(\vec{Q}_z)=S^x(\vec{Q}_x)$. 
All results are obtained on the $N=24$ cluster.}
\end{center}
\end{figure}

\subsection{Selection of stripe orientation by vacancy pairs}

We focus now on the stripe-ordered phase, and will explore how at $ h = 0$,
a pair of defects can, depending on their relative position, favor or disfavor
one of the three stripe directions. In terms of symmetries, this can be
understood as a breaking of $\mathbb{Z}_3$ symmetry related to the spatial
orientation of stripes, while the $\mathbb{Z}_2$ symmetry ($\vec \sigma_i
\rightarrow -\vec \sigma_i$) remains unbroken. Remarkably,
as shown in Fig.~\ref{mstripy_2}, the selected stripe
pattern for a given vacancy pair depends on the coupling constant $\alpha$.\\
This effect is easy to understand for nearest-neighbor defects on a $z-$bond
(configuration $P1$: see Fig.\ref{mstripy_2}); classically, the $z-$stripe
pattern has an energy lower by
\beq
\delta E^z_1 =2-4\alpha
\eeq
than the one of other stripe patterns. Indeed, in the $z$-pattern only one FM 
bond - here a $z$-bond, with a classical contribution to the energy $1-3\alpha$ 
- is lost due to vacancies, instead of two for other patterns (e.g. two 
$x$-bonds, in a $x$-stripe pattern), and for $\alpha>0.5$ FM interactions 
contribute more than AF ones (whose classical contribution to the energy is 
$\alpha-1$). For $\alpha<0.5$ the situation is reversed, which explains that 
$S^z(\vec{Q}_z)$ becomes small and $S^x(\vec{Q}_x)=S^y(\vec{Q}_y) \simeq 1/2$. 

For second neighbors (case $P2$ on Fig.~\ref{mstripy_2}), stripe patterns are
classically degenerate, but this degeneracy is lifted by quantum fluctuations.
We evaluate in second order perturbation theory the contribution to the 
energy of various bonds in which at least one site is a neighbor of the
vacancy; actually only AF bonds give a contribution in this approximation. 
For instance fluctuations on a bond belonging to the same hexagon as the two
impurities yield a contribution $-4(1-2\alpha)^2/[4(1-\alpha)+2(3\alpha-1)]$
to the energy of the $z$-stripe state.
Collecting contributions from all relevant bonds, the energy of
the $z$-stripe state (relative to that of other stripe orientations) is:
\beq
\delta E^z_2=2(1-2\alpha)^2
\big(\frac{1}{5\alpha-1}+\frac{1}{2\alpha}-\frac{2}{1+\alpha} \big),
\eeq
which is positive for $\alpha$ in the vicinity of $1/2$, and changes 
sign at $\alpha_2\simeq 0.651$. This explains the drop of the $z$-stripe
structure factor (dotted curve on Fig.~\ref{mstripy_2}(a)), from large values
for $\alpha \in [0.64(2);0.8]$, to almost zero for smaller $\alpha$.
The smaller values of $S^x(\vec{Q}_x)\simeq 0.3$ in the vicinity of
$\alpha=0.5$ are related to the smaller energy scale separating the
different stripe patterns: the above expression, with its prefactor
$(1-2\alpha)^2$, indicates that the 3 patterns become degenerate at
$\alpha=0.5$. 

A similar energy comparison between different patterns for third neighbor
vacancy pairs (configuration $P3$: see Fig.~\ref{mstripy_2}) predicts a 
change of stripe orientation at $\alpha_3=0.6$ between $z-$stripes 
($\alpha_3<\alpha<\alpha_{LS}$) and $x$-stripes 
($\alpha_{SN}<\alpha<\alpha_3$), following from the perturbative result:
\beq
\delta E^z_3 = \frac{(1-2\alpha)^2(3-5\alpha)}{1-\alpha^2},
\eeq
whereas this orientational change is seen on $N=24$ at a slightly smaller
value $\alpha_3 \simeq 0.53(1)$. Notice that here (and more generally for
pairs of vacancies at further neighbor sites which are not related to each 
other by any reflection symmetry of the honeycomb lattice) the three stripe 
orientations are all non-equivalent to each other.

Finally we note that at zero-field, the conclusions above hold also for pairs
of neighbor vacancies with other relative positions: for instance with
next-nearest neighbor vacancies separated by a $x$- and a $z$-bond and
$\alpha>\alpha_2$, according to the previous reasoning the favored pattern is
a $y$-stripe pattern. 

In presence of a small magnetic field, the $\mathbb{Z}_2$ symmetry
characterizing each stripe orientation (e.g., $z$-stripes) is broken. Moreover,
similar effects occur as in the previously discussed single vacancy case: if
the most favorable pattern in an infinitesimal $[001]$ field (e.g., $z$-stripe 
pattern around a vacancy pair in configuration $P1$, for $\alpha>0.5$) is
untiltable for this field direction, as the field amplitude $h_z$ gets larger
than a critical value $h_{an2}$ this pattern becomes less favorable than other 
patterns which are tiltable. In the limit of large systems $N \gg 1$, a 
classical estimate of $h_{an2}$ gives a result
$4(2\alpha-1)(1-\alpha)/\sqrt{N}$ which, for $\alpha=0.7$ and $N=24$, has a
value $h_{an2} \simeq 0.28$ comparable to the observed position of the
magnetization jump.   

The relevant quantity for the discussion of either low temperature properties or
magnetic ordering at small vacancy concentrations is the energy scale 
$\delta E^z_n$, namely the energy which favors a particular stripe 
pattern in the vicinity of a vacancy pair -- $n$ standing for the number 
of bonds separating both vacancies. Far away from the stripe/liquid transition
(say, for $\alpha \leq 0.7$) $\delta E^z_n$ decreases rapidly with $n$ (e.g.,
$|\delta E^z_n|\leq |\delta E^z_1|/2$ for $n \ge 2$). Closer to
$\alpha_{S/L}$, larger quantum fluctuations make the stripe-orientation
selection by $3^{rd}$ neighbor vacancy pairs as efficient as by nearest
neighbor vacancy pairs. 

Several consequences can result from this: (i) In the simpler case with only 
two vacancies, the energy lowering of some stripe patterns for 
$\alpha \simeq \alpha_{S/L}$ also stabilizes these patterns with
respect to the spin liquid, which leads to a slight shift of the stripe/liquid
transition towards larger $\alpha$ -- possibly $0.82$ to $0.85$ depending on
the relative position of vacancies, as can be seen in Fig.~\ref{mstripy_2}(a). 
(ii) More importantly, if one focuses on systems with substantial 
vacancy concentrations, say, $n_v \sim 10\%$, neighboring vacancy pairs 
as described in Fig.~\ref{mstripy_2} are abundant enough to select locally 
stripe orientations that will frustrate each other at the global scale. 
(iii) As a result, the system could display a glassy
behavior, especially in the vicinity of the stripe-liquid transition. The
resulting glassy phase would consist of stripe-ordered nanoscale domains, with
a vacancy pair in each domain, selecting its stripe orientation.
In order to perform a modeling of such a system and to reproduce a possible 
spin-glass behavior, one should obviously need to consider systems
of (at least) hundreds of Ir atoms, and to take into account the energy cost
of domain walls between regions of different stripe orientations. Moreover,
obviously not all vacancies are grouped into distinct first, second or third 
nearest neighbor vacancy pairs; in particular the effect of unpaired
vacancies, even at zero external field, could be subtle, although one can
imagine single vacancies pinning walls between domains of different stripe
orientation. 

\section{\label{scr} Summary and concluding remarks}

We have considered here several important aspects of the Kitaev-Heisenberg
model, which, despite its simple structure, displays in its $T=0$ phase
diagram three distinct phases, two of which are antiferromagnetically ordered
while the third one is a spin liquid. We employed numerical methods 
(exact diagonalization) for our investigation of the properties
of the model in a magnetic field and the effect of vacancies on the magnetic
properties. The numerical methods were complemented by classical and
perturbative analysis particularly in the frame of the stripe phase.

We showed how the ground state phases evolve in a magnetic field oriented 
along the $[001]$ direction (one of the local easy axes of the
Kitaev-Heisenberg 
Hamiltonian). We found that the three phases are robust to small fields.
Next, we analyzed the influence of the field direction. The magnetic response 
is qualitatively distinct in the different phases. In the stripe 
phase, the magnetization can be described by a spin canting mechanism, 
which also provides an understanding of the anisotropy in the numerically 
obtained susceptibilities. In the liquid phase, we compared the 
effect of small magnetic fields and of small Heisenberg interactions, each 
considered as a perturbation of the Kitaev Hamiltonian. Such perturbations 
are shown to induce specific patterns in the spin-spin correlations beyond 
nearest neighbors, while at larger fields a transition into the 
spin-polarized phase occurs. We determined the phase diagram of the
model in a $[001]$-field, with  good agreement between numerical results and
analytical estimates. For the latter analysis, we took into account both the
spin canting and the quantum fluctuations in the ordered and polarized phases.
This phase diagram has similarities with its counterpart for a $[111]$ field
orientation \cite{JGQT}, however, a major difference between them is that  for
a $[001]$-field we do not see any quantum
critical point in the vicinity of the stripe-liquid phase transition. We
presume that this is related to the absence of a field-induced gap in the
low-energy spectrum of the liquid phase for a $[001]$-field\cite{Kit}.

Subsequently, we focused on the effect of non-magnetic vacancies on the spin
correlations and the magnetic response of the model. For a single vacancy in 
the liquid phase, we found that the spatially anisotropic magnetization
pattern around the vacancy is related to the anisotropy
of spin correlations in the vacancy-free system. Up to finite-size
effects inherent to our approach, these results are also in agreement with
predictions of a non-linear response to a small $[001]$-field, whose effect is
mostly seen on one specific neighbor of the vacancy. In the stripe phase,
a vacancy coupled to a small [001]-field breaks both the $\mathbb{Z}_2$ (time
reversal) and the $\mathbb{Z}_3$ (orientational) symmetry
characterizing the ground state manifold; a single stripe pattern is selected,
which is not continuously tiltable by the field in contrast to the vacancy-free
situation. 

Having in mind the description of systems at finite vacancy concentration,
we considered the interplay of two vacancies sitting either on
nearest or further neighbor sites. Again the magnetic response to an easy-axis 
field depends strongly on the nature of the phase, but it also
depends sensitively on the relative position of vacancies. 
In the liquid phase a non-linear behavior, much stronger than for the
single-impurity case, is seen if the two  vacancies are on the
same sublattice, while the response remains linear if they reside on
different sublattices. In the stripe phase, vacancy pairs select a
specific stripe orientation even at zero field. This selection mechanism,
depending on the value of coupling parameter $\alpha$, can be well understood 
by a perturbative approach evaluating locally the effect of quantum
fluctuations. The response to a small field depends on whether the selected
stripe pattern is tiltable or not for this field direction\cite{tilt}.
Although the strongest selection effect is caused by nearest neighbor
vacancy pairs for a wide range of $\alpha$, further neighbor pairs
have to be considered as well in the vicinity
of the stripe-liquid transition. Therefore we conclude that the intrinsic
randomness of vacancies, in combination with the selection mechanism 
induced by vacancy pairs, may lead to spin-glass behavior in the stripe regime.
There, a vacancy concentration of a few
percents should be enough for the system to show a glassy behavior, with a
tendency to the formation of stripe-ordered nanodomains around vacancy pairs.

Finally, we comment on the possible relevance of this work to layered iridates,
which motivated the derivation of the Kitaev-Heisenberg model. The 
experimental data (magnetic susceptibility, heat capacity\cite{Tak,SG}) on
Na$_2$IrO$_3$ indicate antiferromagnetism at
low temperature and a substantial magnetic anisotropy. Yet, the 
angular dependence of the measured susceptibility is qualitatively opposite 
to that found for a vacancy-free model. We then showed that non-magnetic
vacancies in the dilute limit, i.e., where interactions between vacancies
are irrelevant, can reverse locally this tendency by blocking the spin-canting 
for some field directions. The recent x-ray magnetic scattering 
data \cite{Liu11} ruled out the N\'eel state, implying that the Kitaev 
interaction in Na$_2$IrO$_3$ may dominate over a simple Heisenberg coupling. 
In addition to the stripe phase intrinsic to the Kitaev-Heisenberg model, the 
so-called ''zig-zag'' spin order is also consistent with the data \cite{Liu11}. 
An element in favor of a description of layered iridates A$_2$IrO$_3$ 
(A$=$Li,Na) by the Kitaev-Heisenberg model is the comparison of the low 
temperature susceptibilities of these compounds\cite{SMG}. Using our result 
$\chi \propto \frac{1}{1-\alpha}$ for the stripe phase, we found a relation 
between $\alpha$ values in these two compounds, which indicates that the Li
compound, as the Na-based one, is most probably in the stripe phase but much 
closer to the transition towards the spin liquid. This conclusion is in 
agreement with previous work\cite{SMG,RTT} estimating the relative $\alpha$ 
values from a comparison of the paramagnetic Curie temperatures. 
The application of uniaxial pressure on a single crystal of Li$_2$IrO$_3$ 
might even drive the system into the liquid phase.

To describe the experimental situation in layered iridates on a quantitative 
level, other factors such as lattice distortions have to be considered as 
well. In particular, the lattice distortions may bring about substantial 
anisotropy in the ground state $g$-factors and in the Van Vleck contribution 
to the magnetic susceptibility, originating from transitions to higher lying 
spin-orbital quartet split by non-cubic (tetragonal and/or trigonal) 
crystal fields. Furthermore, these distortions may lead to spin interactions 
not included in the Kitaev-Heisenberg model. For intermediate strength 
of spin-orbit coupling like in iridates, these effects might be essential 
for a quantitative description of magnetic properties. Moreover, our study
suggests that a glassy behavior could occur in the vicinity of the spin liquid
phase; in this context, considering the present model at larger scales, with a
finite density of vacancies, could help to explore the effect of temperature.

\acknowledgments
The authors are indebted to R. Moessner and G. Jackeli for fruitful
discussions at different stages of this work.


\begin{thebibliography}{99}
\bibitem{VVC} A quantitative comparison of our calculated susceptibilities 
to those measured in Ref.~\onlinecite{SG} would include also the effect of 
temperature, especially as the measured anisotropy was also large in the 
paramagnetic phase.
\end{thebibliography}
\end{document}